\documentclass[11pt]{article}
\def\oe{{$\{ p_i \,,\, \rho_i \}\;$}}
\def\of{{$\;\{ p_i \,,\, \phi_i \}\;$}}
\def\oo{{$\;\{ p_i \,,\, \mathcal{E}(\rho_i) \,=\, \widetilde{\rho_i} \}\;$}}

\def\ab{\big | a \big \rangle \big \langle b \big |}
\def \bmatrix#1{ \left [ \matrix{#1} \right ] } 
\def \cmatrix#1{ \matrix{#1} } 

\newcommand{\beq}{\begin{equation}}
\newcommand{\eeq}{\end{equation}}
\newcommand{\beqa}{\begin{eqnarray}}
\newcommand{\eeqa}{\end{eqnarray}}

\begin{document}
\begin{center}
{\Large\bf The Holevo-Schumacher-Westmoreland Channel Capacity
for a Class of Qudit Unital Channels}\\
\bigskip
{\normalsize John Cortese}\\
\bigskip
{\small\it Institute for Quantum Information\\
Physics Department, 
California Institute of Technology 103-33,\\ 
Pasadena, CA 91125 U.S.A.}
\\[4mm]
\date{today}
\end{center}

\begin{center}
\today 
\end{center}

\setlength{\parindent}{0in}
\setlength{\parskip}{0.17in}
\renewcommand{\theequation}{\Roman{equation}}
\begin{abstract}
\noindent
Using the unique nature of the average output state of 
an optimal signalling ensemble,
we prove that for a special class of
$qudit$ unital channels, 
the HSW channel capacity is 
$\mathcal{C} \;=\; \log_2(d) \,-\,min_{\rho} 
\mathcal{S}\left ( \mathcal{E}(\rho)\right )$, where $d$ 
is the dimension of the qudit. 
The result is extended to products of the same class of unital 
qudit channels. 
Thus, the connection between the minimum von Neumann entropy 
at the channel output and the transmission 
rate for classical information over quantum channels extends beyond the 
qubit domain. 
 
\end{abstract}

\section{Introduction}

The Holevo-Schumacher-Westmoreland theorem tells us the asymptotic 
rate at which classical information can be transmitted 
over a quantum channel $\mathcal{E}$ 
per channel use 
is given by the maximum output Holevo 
quantity $\chi$ across all possible signalling ensembles.
$$
\mathcal{C} \;=\; max_{ \{p_i,\rho_i\}} 
\qquad \chi \Big ( \Big \{ p_i , \widetilde{\rho_i } 
\,=\, \mathcal{E}(\rho_i ) \Big \} \Big) 
$$
Here $\chi$ is the Holevo quantity of an ensemble 
$\left \{p_i,\rho_i \right \}$, defined as 
$$
\chi \;=\; 
\mathcal{S}\left ( \sum_i p_i \, \rho_i \right ) \;-\; \sum_i \, p_i 
\, \mathcal{S}\left ( \rho_i \right ) 
$$
where $\mathcal{S}$ is the von Neumann entropy.
\footnote{We shall use $\rho$ to denote a density operator at the channel
input, and $\widetilde{\rho}$ as the corresponding channel output 
density operator. }
We call $\mathcal{C}$ the Holevo-Schumacher-Westmoreland (HSW) channel
capacity. 

We call any input ensemble $\left \{ p_i \,,\, \rho_i \right \}$ that achieves 
$\mathcal{C}$ an optimal ensemble. 
There may be several different optimal input ensembles which achieve 
the optimum HSW channel capacity $\mathcal{C}$.
However, it was shown in \cite{cortese}
that the average channel output state of an optimal ensemble is 
a unique state for all optimal ensembles for that channel.
That is, given a set of optimal input ensembles 
$\left \{ p_i^{(1)} , \rho_i^{(1)} \right \} \;,\; 
\left\{ p_i^{(2)} , \rho_i^{(2)} \right \} \;,\; \cdots \;,\; 
\left\{ p_i^{(N)} , \rho_i^{(N)} \right \}$, 
all achieving $\mathcal{C}$, 
we define 
$\widetilde{\Phi}^{(k)} \;=\; 
\mathcal{E}\left (\sum_i p_i^{(k)} \, \rho_i^{(k)} \right )$.
Then it has been shown we must have 
$\widetilde{\Phi}^{(1)} \;=\; \widetilde{\Phi}^{(2)} 
\;=\; \cdots \;=\; \widetilde{\Phi}^{(N)}$.

The main idea of this paper is the unique nature of the 
output ensemble average state of an optimal signalling ensemble
for a quantum channel $\mathcal{E}$
tells us alot about $\mathcal{C}$ for that channel.

\section{Background Material} 

\subsection{Invariance of $\mathcal{S}$ and $\chi$ under unitary operators}

Consider any ensemble 
$\left\{ p_i , \rho_i \right \}$. 
Acting on each $\rho_i$ with the same unitary operator $U$ yields a
set of valid quantum states $U \rho_i U^{\dagger}$ and the ensemble
$\left\{ p_i , U \rho_i U^{\dagger} \right \}$. 
Furthermore, each $\rho_i$ has the same eigenvalues as the corresponding 
$U\rho_iU^{\dagger}$. Since von Neumann entropy depends only 
on a density operators eigenvalues, we conclude 
$\mathcal{S}(\rho_i) \;=\; \mathcal{S}\Big ( U\rho_iU^{\dagger}\Big)$.
Furthermore, this implies the Holevo quantity $\chi$ of the ensembles 
$\left\{ p_i , \rho_i \right \}$ and 
$\left\{ p_i , U \rho_i U^{\dagger} \right \}$ is equal, since
\begin{equation}\label{E:1}
\chi \left ( \left \{ p_i , U \rho _i U^{\dagger} \right \} \right ) \;=\;  
\mathcal{S} \left ( \sum_i \, p_i \,  U \rho _i U^{\dagger} \right ) \;-\;  
\sum_i  \, p_i \, \mathcal{S} \left ( U \rho _i U^{\dagger} \right ) 
\end{equation}
$$
\;=\; 
\mathcal{S} \left ( U \left ( \sum_i \, p_i \,  \rho _i \right )  
U^{\dagger} \right ) \;-\;  
\sum_i  \, p_i \, \mathcal{S} \left ( U \rho _i U^{\dagger} \right ) 
$$
$$
\;=\; 
\mathcal{S} \left ( \sum_i \, p_i \,  \rho _i \right ) \;-\;  
\sum_i  \, p_i \, \mathcal{S} \left ( \rho _i  \right ) \;=\; 
\chi \Big ( \Big \{ p_i , \rho _i \Big \} \Big ) .
$$

\section{HSW Channel Capacity for single qubit unital channels}

As an example of the approach we shall be taking, we derive the HSW channel
capacity for single qubit unital channels. This result was previously 
derived in \cite{Ruskai99a} by a different technique. 

We describe a single qubit density operator using the Bloch sphere 
representation.
$$
\rho \; = \;  \frac{1}{2} \; \left ( \; \mathcal{I} \; + \; 
\vec{\mathcal{W}}_{\rho}
\bullet \vec{\sigma} \; \right )
$$
The symbol $ \vec{\sigma}$ is the vector of 2 by 2 Pauli matrices 
$$  \vec{\sigma} \; = \; 
\bmatrix{ \sigma_x \cr \sigma_y \cr \sigma_z }   
\;\;\;\;\; where 
\;\;\;\;\; \sigma_x \;=\; \bmatrix{ 0  & 1 \cr 1  & 0 },
\;\;\;\;\; \sigma_y \;=\; \bmatrix{ 0  & -i \cr i & 0 },
\;\;\;\;\; \sigma_z \;=\; \bmatrix{ 1  & 0 \cr 0 & -1 }
\;.
$$
The Bloch vector $\vec{\mathcal{W}}$ is a real,
three dimensional vector
which has magnitude equal to one when representing a pure state density matrix,
and magnitude less than one for a mixed (non-pure) density matrix.  

It was shown in \cite{Ruskai99a} that the action of 
a single qubit unital channel $\mathcal{E}$ 
on an input state $\rho$ could be represented as 
$\widetilde{\rho}\;=\;  \mathcal{E}(\rho)$, where $\rho$ has Bloch vector
$\bmatrix{ w_x  \cr w_y  \cr w_z }$ 
and 
$\widetilde{\rho}$
has Bloch vector
$\bmatrix{ \lambda_x \, w_x  \cr \lambda_y \, w_y  \cr \lambda_z \, w_z }$. 
Here the $\lambda_k \in [-1,1]$. Using the unique nature of the average 
output state of an optimal signalling ensemble, we shall show the 
HSW channel capacity $\mathcal{C}$ is 
$\mathcal{C} \;=\; 1 \,-\, max_{\rho} \; \mathcal{S}(\mathcal{E}(\rho))$.

\subsection{Achievability of Output Ensembles}

We say an ensemble $\{q_j\,,\, \phi_j \}$ at the channel output is 
{\it achievable }
if there exists an input ensemble $\{ q_j\,,\, \varphi_j\}$ such that
the $\{\varphi_j\}$ are all valid density operators and 
$\mathcal{E}(\varphi_j) \;=\; \phi_j \; \; \forall j$.
Let us recall some properties of the Pauli matrices 
$\{ \sigma_k\}$. The $\{ \sigma_k\}$ obey the relations 
$\sigma_i \sigma_j \;=\; - \, \sigma_j \sigma_i \; for \; i\ne j $
and 
$\sigma_i \sigma_j \;=\; I_2\;for \; i =  j $. 
\footnote{We write $I_d$ for the $d$ by $d$ identity matrix.} 
Thus, we find 
$\sigma_i \sigma_j \sigma_i \;=\; - \, \sigma_j \; for \; i\ne j $
and 
$\sigma_i \sigma_j \sigma_i \;=\; \sigma_i\;for \; i =  j $. 
The $\sigma_k$ are Hermitian, so $\sigma_k^2 \;=\; I_2$ implies the 
$\sigma_k$ are unitary, yielding $\sigma_k^{\dagger} \,=\, \sigma_k$. 

Let \oe be an optimal input ensemble with corresponding
output ensemble \oo. Apply a Pauli operator $\sigma_k$ to all the 
density matrices in \oo, yielding an ensemble 
$\{ p_i\,,\, \sigma_k \, \widetilde{\rho_i} \, \sigma_k^{\dagger} \}$.
We know the density operators 
$\{\sigma_k \, \widetilde{\rho_i} \, \sigma_k^{\dagger} \}$ 
are valid density operators because 
$\sigma_k$ is a unitary operator, and hence acting with
$\sigma_k$ implements a change of basis at the channel 
output. The question we are interested in is whether the output ensemble 
$\{ p_i\,,\, \sigma_k \, \widetilde{\rho_i} \, \sigma_k^{\dagger} \}$ is 
{\it achievable}.
To answer this, we know for each
$\widetilde{\rho_i}$, there is a valid input 
$\rho_i$ such that $\mathcal{E}(\rho_i) \;=\; \widetilde{\rho_i}$.
Consider the following. 
$$
\sigma_k \, \widetilde{\rho_i } \, \sigma_k^{\dagger} \;=\; 
\sigma_k  \, \mathcal{E} \left ( \rho_i \right ) \, \sigma_k^{\dagger}  
\;=\; 
\sigma_k  \; \mathcal{E}\left ( \frac{1}{2}\Big ( I_2   
\,+\, \omega_{x_i}  \sigma_x 
\,+\, \omega_{y_i}  \sigma_y   
\,+\, \omega_{z_i}  \sigma_z \Big ) \right )  \; \sigma_k ^{\dagger}  
$$
$$
\;=\; 
\sigma_k \, \left ( \frac{1}{2}\Big ( I_2   
\,+\, \lambda_x \omega_{x_i} \sigma_x 
\,+\, \lambda_y \omega_{y_i} \sigma_y 
\,+\, \lambda_z \omega_{z_i} \sigma_z \Big )\right ) \, \sigma_k^{\dagger}  
$$
$$
\;=\; 
\frac{1}{2}\Big ( I_2   
\,+\, \lambda_x \omega_{x_i} \sigma_k  \sigma_x \sigma_k ^{\dagger}  
\,+\, \lambda_y \omega_{y_i} \sigma_k  \sigma_y \sigma_k ^{\dagger}  
\,+\, \lambda_z \omega_{z_i} \sigma_k \sigma_z \sigma_k  ^{\dagger}  \Big )
$$
Define $\bar{\delta}_{k,l} \;=\; 0$ if $k =l$, and $1$ if $k\ne l$.
Note that $\sigma_k \, \sigma_l \, \sigma_k \,=\, 
\left ( -1 \right )^{\bar{\delta}_{k,l}} \, \sigma_l$.
If $\varphi_i$ has the Bloch vector 
$\bmatrix{ \left ( -1\right ) ^{\bar{\delta}_{k,x}}\,  \omega_x  \cr  
\left ( -1\right ) ^{\bar{\delta}_{k,y}}\, \omega_y  
\cr  \left ( -1\right ) ^{\bar{\delta}_{k,z}}\, \omega_z }$, 
then the channel output of $\varphi_i$ is
$$
\mathcal{E}(\varphi_i) 
\;=\; 
\frac{1}{2} \, \Big ( \, I_2 
\,+\, 
\left ( -1\right ) ^{\bar{\delta}_{k,x}}\,
\lambda_x \omega_x \sigma_x 
\,+\, 
\left ( -1\right ) ^{\bar{\delta}_{k,y}}\,
\lambda_y \omega_y \sigma_y 
\,+\, 
\left ( -1\right ) ^{\bar{\delta}_{k,z}}\,
\lambda_z \omega_z \sigma_z 
\, \Big )
$$
$$
\;=\;
\frac{1}{2} \, \Big ( \, I_2 
\,+\, 
\lambda_x \omega_x \sigma_k \sigma_x  \sigma_k^{\dagger}
\,+\, 
\lambda_y \omega_y  \sigma_k \sigma_y  \sigma_k^{\dagger}
\,+\, 
\lambda_z \omega_z  \sigma_k \sigma_z  \sigma_k^{\dagger}
\, \Big ) 
\;=\; 
\sigma_k \, \mathcal{E}\left ( \varphi_i \right )  \, \sigma_k^{\dagger}. 
\;=\; 
\sigma_k \widetilde{\rho_i} \sigma_k^{\dagger}. 
$$
If we can show the 
$\varphi_i$ are valid density operators, then we have shown that the 
output ensemble 
$\{ p_i\,,\, \sigma_k \, \widetilde{\rho_i} \, \sigma_k^{\dagger} \}$
is achievable. In order for $\varphi_i$ to be a valid density operator,
we must have 
the corresponding Bloch vector composed of three real entries, and
the magnitude of the Bloch vector less than or equal to one. Since
the $\rho_i$ are valid density operators, the
three $\omega_k$ are real, and obey
$\omega_x^2 \,+\,\omega_y^2 \,+\, \omega_z^2 \;\le 1$. 
Now $\left ( -1\right ) ^{\bar{\delta}_{k,l}} \; for \; k,l=\{x,y,z\}$ is real
and equal in magnitude to one.
The magnitude of the Bloch vector for $\varphi_i$ is
$\Big (\, \left ( -1\right ) ^{\bar{\delta}_{k,x}}\,\omega_x\,\Big )^2 \,+\,
\Big (\,\left ( -1\right ) ^{\bar{\delta}_{k,y}}\,\omega_y\,\Big )^2 \,+\,
\Big (\,\left ( -1\right ) ^{\bar{\delta}_{k,z}}\,\omega_z\,\Big )^2 \;=\; 
\omega_x^2 \,+\, \omega_y^2 \,+\, \omega_z^2 \;\le1$, where the last inequality
follows from our knowledge that the $\rho_i$ are valid density operators.  
Thus the $\varphi_i$ are valid density operators.  We conclude that 
if there exists an optimal input ensemble \oe, with corresponding 
output ensemble \oo, then the ensemble 
$\Big \{p_i \,,\, \sigma_k \widetilde{\rho_i} \sigma_k^{\dagger}\Big \}$ is 
achievable, with corresponding input ensemble $\{p_i\,,\, \varphi_i\}$.
Furthermore, the input ensemble $\{p_i\,,\, \varphi_i\}$ is optimal, since 
$\sigma_k$ is a unitary operator, and we showed in equation (\ref{E:1})
that a unitary
operator acting on an ensemble does not change the Holevo quantity of that
ensemble. Since \oo attained the maximal Holevo quantity $\mathcal{C}$
at the channel output, the output ensemble 
$\Big \{p_i \,,\, \sigma_k \widetilde{\rho_i} \sigma_k^{\dagger}\Big \}$
also has a Holevo value of $\mathcal{C}$. Thus
$\{p_i\,,\, \varphi_i\}$ is an optimal input ensemble.

To summarize, we first 
chose a basis of operators $E_i$, in this case 
the identity $I_2$ and the three Pauli operators $\{ 
\sigma_x \,,\,  
\sigma_y \,,\,  
\sigma_z \}$, 
in which to expand the density matrix
$\rho\;=\; \sum_i \, \alpha_i \, E_i$. Next, we found a set of 
unitary operators $U_k$, in this case again the Pauli operators $\sigma_k$, 
such that the $U_k$ act on the $E_i$ resulting in a multiplicative 
phase factor :
$U_k E_i U_k^{\dagger} \;=\; \kappa_{(k,i)}\, E_i$, where 
$\kappa_{(k,i)}$ is a complex 
quantity.
The unital nature of the qubit channel $\mathcal{E}$ tells us 
that $\mathcal{E}(E_i) \,=\, \lambda_i \, E_i \quad \forall \; i\;$ in the
operator basis $\{E_i\}$. This leads to the commutation of the channel 
$\mathcal{E}$ with the set of unitaries $\{U_k\}\,=\, \big \{ \pm I_2 
\,,\,\pm \sigma_x  
\,,\,\pm \sigma_y 
\,,\,\pm \sigma_z \big \}$.   
$$
U_k \mathcal{E}\left (E_i\right )U_k^{\dagger} \,=\, 
U_k \lambda_i E_i U_k^{\dagger} \,=\, \lambda_i U_k E_i U_k^{\dagger} \,=\, 
\lambda_i \, \kappa_{(k,i)}\, E_i \,=\, 
\kappa_{(k,i)}\, \mathcal{E}\left(E_i\right) 
$$
$$
( \; By \; linearity \, of \; quantum \; channels \; ) \; \,=\, 
\mathcal{E}\left(\kappa_{(k,i)}\, E_i\right) \,=\, 
\mathcal{E}\left(U_k \, E_i \, U_k^{\dagger}\right). 
$$ 
Since we have an expansion of $\rho$ in terms of the $E_i$, using the 
linearity of quantum channels, we conclude 
that 
\begin{equation}\label{E:2}
U_k \, \mathcal{E}\left(\rho\right)\, U_k^{\dagger} \,=\, 
U_k\, \mathcal{E}\left(\frac{1}{2} \, \sum_i \, \alpha_i \, E_i \right)\, 
U_k^{\dagger} \,=\, 
U_k\, \left( \frac{1}{2} \, 
\sum_i \, \alpha_i \, \mathcal{E}\left ( E_i \right)\, \right ) \, U_k^{\dagger} 
\,=\, \frac{1}{2} \, 
\sum_i \, \alpha_i \, U_k\,\mathcal{E}\left(E_i \right)\,U_k^{\dagger} 
\end{equation}
$$
\,=\, \frac{1}{2} \, 
\sum_i \, \alpha_i \, \mathcal{E}\left(U_k\,E_i \,U_k^{\dagger} \right)
\,=\, 
\mathcal{E}\left(\frac{1}{2} \, \sum_i \, \alpha_i \, U_k\,E_i \,U_k^{\dagger}\right) \,=\, 
\mathcal{E}\left(U_k \left ( \frac{1}{2} \, \sum_i \, \alpha_i E_i \right ) 
U_k^{\dagger}\right) 
\,=\, 
\mathcal{E}\left(U_k\,\rho\,U_k^{\dagger}\right).
$$
A $U_k$ acting at the input is a basis change and hence
$U_k \, \rho \,U_k^{\dagger}$ is a valid input density operator.
Equation (\ref{E:2}) allows us to conclude that any $U_k$ 
acting on the output states $\widetilde{\rho}_i$ 
of an optimal ensemble \oe yields an output ensemble
$\{ p_i \,,\, U_k \,\widetilde{\rho}_i\, U_k^{\dagger} \}$ which is
achievable. 
The achievability of channel output ensembles generated by
$U_k$ acting on the output ensemble of an optimal 
input ensemble will be a critical tool in extending our unital 
qubit channel analysis
to the determination 
of the Holevo-Schumacher-Westmoreland channel capacity $\mathcal{C}$ 
for a special class of {\it qudit} unital channels. 

\subsection{Using Symmetry Properties of Optimal Ensembles} 

Consider a unital qubit channel with an optimal input ensemble 
$\left\{ p_i , \rho_i \right \}$,
\footnote{That such an ensemble exists was shown in \cite{Schumacher99}. } 
average input state $\Phi\;=\; \sum_i \, p_i \rho_i$
and average output state
$\widetilde{\Phi}\;=\; \mathcal{E}(\Phi)$. 
Let $\Phi$ have Bloch vector $\vec{V} \;=\; 
\bmatrix{ v_x  \cr v_y  \cr v_z }$ 
and $\widetilde{\Phi}$ have Bloch vector $\widetilde{\vec{V}} \;=\; 
\bmatrix{ \widetilde{v}_x  \cr \widetilde{v}_y  \cr \widetilde{v}_z } \;=\; 
\bmatrix{ \lambda_x \, v_x  \cr \lambda_y \, v_y  \cr \lambda_z \, v_z }$.  
Choose one of the three $\{\sigma_k\}$ and 
apply this $\sigma_k$
to the output states $\widetilde{\rho}_i$
to obtain a new output ensemble 
$\{p_i\,,\, \sigma_k \widetilde{\rho}_i \sigma_k^{\dagger} \} 
\;\equiv \; \{p_i\,,\, \widetilde{\rho}_i'\} $.  
We know from our work above that the output ensemble  
$\{p_i\,,\, \sigma_k \widetilde{\rho}_i\sigma_k^{\dagger} \} $
is achievable and optimal.
The action of $\sigma_k$ on the output ensemble \oo generates a 
corresponding transformation of the 
average output state of the optimal ensemble
$\widetilde{\Phi}$, 
$$
\sum_i \, p_i\, \sigma_k \,\widetilde{\rho}_i\,
\sigma_k^{\dagger} 
\;=\; 
\sigma_k \, \left ( \, \sum_i \, p_i\,\widetilde{\rho}_i\,\right ) \,
\sigma_k^{\dagger} 
\;=\; \sigma_k \widetilde{\Phi} \sigma_k^{\dagger}
\;=\; 
\widetilde{\Phi}' .
$$
By the invariance property shown in \cite{cortese}, we have
$\widetilde{\Phi}' \;\equiv\; \widetilde{\Phi}$. Now 
$\widetilde{\Phi}$ has Bloch vector $\widetilde{\vec{V}} \;=\; 
\bmatrix{ \widetilde{v}_x  \cr \widetilde{v}_y  \cr \widetilde{v}_z } \;=\; 
\bmatrix{ \lambda_x \, v_x  \cr \lambda_y \, v_y  \cr \lambda_z \, v_z }$
and 
$\widetilde{\Phi}'$ has Bloch vector $\widetilde{\vec{V}}' \;=\; 
\bmatrix{ \left ( -1\right ) ^{\bar{\delta}_{k,x}}\,  \widetilde{v}_x\cr  
\left ( -1\right ) ^{\bar{\delta}_{k,y}}\,  \widetilde{v}_y
\cr  \left ( -1\right ) ^{\bar{\delta}_{k,z}}\,  \widetilde{v}_z}$.
For $k = \{x,y,z\}\;, \;\widetilde{\Phi} \;\equiv \; \widetilde{\Phi}'$ 
implies 
\begin{equation}\label{E:3}
\widetilde{v}_x \;=\; \left ( -1\right ) ^{\bar{\delta}_{k,x}}\,  \widetilde{v}_x
\qquad and \qquad 
\widetilde{v}_y \;=\; \left ( -1\right ) ^{\bar{\delta}_{k,y}}\,  \widetilde{v}_y
\qquad and \qquad 
\widetilde{v}_z \;=\; \left ( -1\right ) ^{\bar{\delta}_{k,z}}\,  \widetilde{v}_z.
\end{equation}
The only way all three relationships in equation (\ref{E:3}) 
can be true $\forall \; k=\{x,y,z\}$ 
is if $\widetilde{v}_x=  \widetilde{v}_y=  \widetilde{v}_z= 0$. 
The fact 
$\widetilde{\Phi}$ has Bloch vector $\widetilde{\vec{V}} \;=\; 
\bmatrix{ 0 \cr  0 \cr  0 }$ leads to the conclusion that 
$\widetilde{\Phi} \,=\, \frac{1}{2} \left ( I_2 \,+\, \widetilde{\vec{V}} 
\bullet \vec{\sigma} \right )  \,=\, \frac{1}{2}\, I_2\; $ for all 
optimal ensembles.  

A second way to see that $\widetilde{\Phi} \;\equiv \; \frac{1}{2} \, I_2$ 
is via Schur's lemma\cite{cornwell}. Consider the group 
$\mathcal{H}$ composed of the 
eight operations 
$\{ \, \pm I_2 \,,\, \pm \sigma_x \,,\, \pm \sigma_y \,,\, \pm \sigma_z \,\}$.
A necessary and sufficient condition for a finite group
$\mathcal{G}$ to be irreducible is if
the relation  $\frac{1}{\| \mathcal{G}\|} 
\sum_{g\in\mathcal{G}} \, \Big \vert Trace[g] \Big | ^2 \;=\; 1$
is true\cite{cornwell}. 
Here $\|\mathcal{G}\|$ is the order of the group $\mathcal{G}$.
Noting our group $\mathcal{H}$ above is finite, and 
computing the sum with our group $\mathcal{H}$, we find 
$\mathcal{H}$ is irreducible. 

Schur's lemma states that if a group $\mathcal{G}$ is irreducible and 
has a $d$ dimensional representation  $\Gamma\left ( \mathcal{G}\right ) $
in which each representation element
$\Gamma(g)$ commutes with a $d$ by $d$ matrix 
M $\forall g \in \mathcal{G}$, then M is proportional to $I_d$\cite{cornwell}. 
The fact that we found 
$\sigma_k \widetilde{\Phi} \sigma_k^{\dagger} \,=\, \widetilde{\Phi}  \quad
\forall k \in \{x,y,z\}$, together with the same trivial result for $I_2$,
implies that all elements of $\mathcal{H}$ commute with $\widetilde{\Phi}$ 
and thus $\widetilde{\Phi} \;\propto\; I_2$. The trace condition
$Trace\left (\widetilde{\Phi}\right ) \,=\, 1$ leads us to conclude 
$\widetilde{\Phi} \,=\, \frac{1}{2} \, I_2$. 

Having determined 
$\widetilde{\Phi}$, we can now rewrite the 
Holevo-Schumacher-Westmoreland channel
capacity $\mathcal{C}$ as 
$\mathcal{C}\;=\; \log_2(2) \,-\, \sum_i p_i \, 
\mathcal{S}\left ( \mathcal{E} \left ( \rho_i \right ) \right ) $.
To further simplify this result, we use two results from \cite{Schumacher99}.
In their paper, 
Schumacher and Westmoreland 
worked with the relative entropy function, $\mathcal{D}\left [ \rho \| \phi \right ]$ defined as $Trace\left [ \rho \log_2 (\rho) \,-\, \rho \log_2(\phi) \right ]$.
Using $\mathcal{D}$, they 
proved the following two results.

\vspace{0.2in}

I) {\it The equal distance property of optimal ensembles. }
\newline
For any optimal ensemble \oe, we have 
\begin{equation}\label{E:4}
\mathcal{D}\Big [ \mathcal{E}(\rho_i ) \Big \| \mathcal{E}(\Phi) \Big ] 
\;=\; 
\mathcal{C}
\quad \forall i. 
\end{equation}

II) {\it The sufficiency of the maximal distance property. }
\newline 
For any optimal ensemble \oe with average input state
$\Phi \;=\; \sum_i p_i \rho_i$, we have 
\begin{equation}\label{E:5}
\mathcal{D}\Big [ \mathcal{E}(\phi) \Big \| \mathcal{E}(\Phi) \Big ] 
\; \le \; \mathcal{C}\quad for \quad any \quad input 
\quad density \quad matrix \quad \phi. 
\end{equation}

\noindent
In both I) and II), $\Phi \;=\; \sum_i p_i \rho_i$ and 
$\mathcal{C}$ is the Holevo-Schumacher-Westmoreland channel capacity.  
For the case of qubit unital channels, we have found that 
every optimal ensemble \oe must obey $\mathcal{E}\Big (
\sum_i p_i \, \rho_i \Big ) \;=\; \frac{1}{2} I_2$.  
Looking at the relative entropy formula, we see that
$\mathcal{D}\Big[ \mathcal{E}\left ( \phi\right ) \Big \| \frac{1}{d} \, 
I_d \Big ] \;=\; \log_2(d) \,-\, \mathcal{S}\left ( \mathcal{E}
\left ( \phi\right) \right )$, where $\mathcal{S}$ is the von Neumann 
entropy and $\phi$ is any input density matrix. 
(For more details on this derivation, please see 
the appendices in \cite{cortese}.) 
Using the fact that for qubit unital channels we have 
found, for all optimal ensembles \oe, 
that 
$\mathcal{E}\Big ( \sum_i p_i \, \rho_i \Big ) \,=\, \frac{1}{2} \,I_2$,
the above two Schumacher and Westmoreland results become, in the qubit 
unital channel case, 

\vspace{0.2in}

I') \begin{equation}\label{E:6} 
1 \,-\, \mathcal{S}\left ( \mathcal{E}( \rho_i ) \right ) \;= \; 
\mathcal{C} \quad \forall i \qquad implying \quad 
\mathcal{S}\left( \mathcal{E}\left ( \rho_i \right ) \right ) \,=\, 
\mathcal{S}\left( \mathcal{E}\left ( \rho_j \right ) \right ) \;\forall \; i,j.  
\end{equation}

II') 
\begin{equation}\label{E:7} 
1 \,-\, \mathcal{S}\left ( \mathcal{E}( \phi ) \right ) \;\le \; 
\mathcal{C} \quad \forall \quad input \quad density \quad matrices \quad \phi .
\end{equation}

\vspace{0.2in} 

We know that II') is achieved with equality when $\phi$ is any of the
$\rho_i$ in the optimal ensemble \oe. Thus I') and II') 
taken together yield 
$1\,-\, \mathcal{S}\left ( \mathcal{E}(\phi)\right) \;\le \; 
1\,-\, \mathcal{S}\left ( \mathcal{E}(\rho_i)\right)$ or
$\mathcal{S}\left ( \mathcal{E}(\phi)\right) \;\ge \; 
\mathcal{S}\left ( \mathcal{E}(\rho_i)\right)$,
which, since $\phi$ can be any input density matrix, implies
$ \mathcal{S}\left ( \mathcal{E}( \rho_i ) \right )  \;=\; 
min_{\phi} \; \mathcal{S}\left ( \mathcal{E}( \phi ) \right ) $. 
Plugging this result into I') yields our final result for 
the Holevo-Schumacher-Westmoreland channel capacity for qubit unital channels. 
$$
\mathcal{C} \;=\; 1 \,-\, 
min_{\phi} \, \mathcal{S}\left ( \mathcal{E}(\phi) \right ) . 
$$
For qubit unital channels, the minimum channel output von Neumann 
entropy determines the Holevo-Schumacher-Westmoreland 
channel capacity $\mathcal{C}$. 

\subsection{Ensemble Achievability}

The achievability of a transformed output ensemble 
is a concept worth emphasizing. 
In our discussion of unital qubit channels, the reason 
why we could conclude the average output state of an
optimal ensemble commuted with all eight members of our group 
$\mathcal{H} \;=\; \{ 
\pm I_2 \, , 
\pm \sigma_x \, , 
\pm \sigma_y \, , 
\pm \sigma_z \}$ 
was because, given an optimal 
ensemble \oe, each of the eight output ensembles 
$\{ p_i \,,\, h \widetilde{\rho_i} h^{-1} \}$, where
$h \in \mathcal{H}$, was achievable. 
The existence of an optimal input ensemble \of which maps 
via the quantum channel $\mathcal{E}$ to 
$\{ p_i \,,\, h \widetilde{\rho_i} h^{-1} \}$ 
is what allowed us to conclude 
the relationship $h \widetilde{\Phi} h^{-1} \;=\; \widetilde{\Phi}$ 
was valid, and apply Schur's lemma.

For a generic group $\mathcal{M}$ acting on the channel output 
of an optimal ensemble \oe, there will typically 
be $m_0\, \in \mathcal{M}$ such that 
$\Big \{ p_i \,,\, m_0 \widetilde{\rho_i} m_0^{-1} \Big \}$ are not 
achievable ensembles. In these cases, we cannot conclude 
$m_0 \widetilde{\Phi} m_0^{-1} \;=\; \widetilde{\Phi}$ holds, 
where $\widetilde{\Phi}$ is the average output state of 
an optimal ensemble. Yet it was the fact that 
$m_0 \widetilde{\Phi} m_0^{-1} \;=\; \widetilde{\Phi}$
holds $\forall m \in \mathcal{M}$ that led us to 
apply Schur's Lemma and conclude $\widetilde{\Phi} \;\propto I_2$.
The lack of achievability for one or more of the transformed output ensembles
$\Big \{ p_i , m \, \widetilde{\rho}_i \, m^{-1} \Big \}$
prevents us from appealing to Schur's Lemma. 
An example of the limitations to determining HSW channel 
capacity which results from output ensemble
non-achievability arises in the case of 
non-unital qubit channels.

\subsection{A Non-Unital Qubit Channel Example}

Our technique fails for non-unital qubit channels. The reason why is 
the lack of achievability of output 
ensembles generated by members of the Pauli group acting
on an output 
optimal ensemble. For example,
consider the non-unital linear qubit channel specified in the 
Ruskai-King-Swarez-Werner notation as 
$\{ t_x = t_y =0, t_z =0.2, \lambda_x=\lambda_y=0,\lambda_z = 0.4\}$.
This channel maps an input Bloch vector $\vec{W}$ to an output Bloch
vector $\widetilde{\vec{W}}$ as :

$$
\vec{W} \;=\; \bmatrix{ w_x \cr w_y \cr w_z } \; \rightarrow  \;
\bmatrix{ 0 \cr 0 \cr t_z \,+\, \lambda_z w_z } \;=\; 
\bmatrix{ 0 \cr 0 \cr 0.2 \,+\, 0.4  w_z } \;=\; \widetilde{\vec{W}} .
$$

By inspection, an optimal input ensemble is \oe
with $\rho_{1,2} \;=\; \frac{1}{2} \left ( I_2 \;\pm\; \sigma_z \right )$,
and corresponding output density matrices 
$\widetilde{\rho_1} \;=\; \frac{1}{2} \left ( I_2 \; -\; 0.2 \sigma_z\right )$
and  
$\widetilde{\rho_2} \;=\; \frac{1}{2} \left ( I_2 \; +\; 0.6 \sigma_z\right )$.
Numerical analysis for this channel indicates 
the optimum output average state is 
$\widetilde{\Phi} \;\approx \; \frac{1}{2} \left ( I_2 \;+\; 0.2125 \, \sigma_z\right)$. 
Since 
$\widetilde{\Phi} \,\ne \, \frac{1}{2} \, I_2 $, we anticipate we will 
not be able to meet the conditions for the application of Schur's lemma. 

Consider applying the unitary operator $\sigma_z$
to the output optimal ensemble \oo determined in the previous paragraph.
We obtain 
$$
\sigma_z \, \widetilde{\rho_1} \, \sigma_z \,=\,
\sigma_z \, \bigg ( \frac{1}{2} \, \Big ( I_2 \,-\, 0.2 \sigma_z \Big ) \, \bigg ) 
\, \sigma_z^{\dagger} 
\,=\, \widetilde{\rho_1}
\qquad and \qquad
\sigma_z \, \widetilde{\rho_2} \, \sigma_z \,=\,
\sigma_z \, \bigg ( \frac{1}{2} \, \Big ( I_2 \,+\, 0.6 \sigma_z \Big ) \, \bigg ) 
\, \sigma_z^{\dagger} 
\,=\, \widetilde{\rho_2}.
$$
Thus the output ensemble $\Big \{ p_i \,,\, \sigma_z \, 
\mathcal{E}\left ( \rho_i \right ) \, \sigma_z^{\dagger} \,=\, 
\sigma_z \,  \widetilde{\rho}_i \, \sigma_z^{\dagger} \Big \}$
is identical to the output ensemble 
$\Big \{ p_i \,,\,
\mathcal{E}\left ( \rho_i \right ) \,=\, \widetilde{\rho}_i \, \Big \}$,
both being generated by the input 
ensemble \oe. Thus 
the output ensemble $\Big \{ p_i \,,\, \sigma_z \, 
\mathcal{E}\left ( \rho_i \right ) \, \sigma_z^{\dagger} \,=\, 
\sigma_z \,  \widetilde{\rho}_i \, \sigma_z^{\dagger} \Big \}$
is an achievable output ensemble. 

The 
application of $\sigma_x$ or $\sigma_y$ to \oo however does not yield
an achievable ensemble. 
To see why, consider applying $\sigma_x$ to 
$\widetilde{\rho_2} \;=\; \frac{1}{2} \left ( I_2 \; +\; 0.6 \sigma_z\right )$,
which since $\sigma_x \sigma_z \sigma_x^{\dagger} \,=\, -\sigma_z$, 
yields the output density operator  
$\widetilde{\rho_2}' \;=\; \frac{1}{2} \left ( I_2 \; -\; 0.6 \sigma_z\right )$.
The corresponding input density operator would have Bloch vector
$\vec{W}' \;=\; \bmatrix{ \phantom{-}0 \cr \phantom{-}0 \cr -2 } $,
which is not a valid qubit density operator, since
$\big \| \vec{W}'\big\| \,> \, 1$. Since the output state 
$\sigma_x \widetilde{\rho_2} \sigma_x^{\dagger}$ can never be mapped 
to by a valid input qubit density operator, 
we cannot assume the relation $\sigma_x \, \widetilde{\Phi} \,
\sigma_x^{\dagger} \;=\; \widetilde{\Phi}$ holds.
Thus, we do not have the necessary Schur commutation requirement 
that $g \widetilde{\Phi} \;=\; \widetilde{\Phi} g$ for all members 
$g$ of the Pauli group $\{\pm I_2,\pm \sigma_x,\pm \sigma_y,\pm \sigma_z\}$, 
and hence cannot conclude $\widetilde{\Phi} \;=\; \frac{1}{2} I_2$, 
as we anticipated.

As we shall develop in more detail below,
working with qudits, if we can find a group
$\mathcal{G}$ such that we are assured all elements $g\in\mathcal{G}$ 
are unitary and acting on the output states of 
an optimal ensemble \oe yield achievable ensembles 
$\forall \, g\in \mathcal{G}$,
than we will be able to conclude the average output
state of any optimal ensemble is $\widetilde{\Phi} \;=\; 
\frac{1}{d} \, I_d$. From this conclusion, we can use 
the Schumacher-Westmoreland relative entropy results from equations
\ref{E:4},\ref{E:5},\ref{E:6}, and \ref{E:7} to conclude the 
states in any input
optimal ensemble must be a subset of those input states 
which yield the minimum output von Neumann entropy. 
This in turn leads us to a HSW channel capacity $\mathcal{C}$ of
$$
\mathcal{C} \;=\; \log_2(d) \;-\; 
min_{\phi} \, \mathcal{S}\left ( \mathcal{E}(\phi) \right ) 
$$
for those qudit channels to which we can successfully apply Schur's
lemma. We now proceed to determine the subset of
qudit channels which meet the Schur's lemma requirements. 

\section{Qudit Channels}

The HSW channel capacity result 
for unital qubit channels was
previously proven in \cite{Ruskai99a} by a method which did
not generalize to the general qudit case (ie: for qudit dimension d $>$ 2).
The technique discussed 
in this paper does generalize to a special subclass of unital qudit 
channels. Before describing that generalization, we present
some background material on qudits and qudit channels. 

\subsection{Qudits}

A qudit is a system with $d$ orthogonal 
pure states 
$\big | j \big \rangle, \, j =0,1,2,\cdots, d-1$.
The generalization of the qubit Pauli operators $\sigma_x$ and $\sigma_z$
are the two operators $\hat{X}$ and $\hat{Z}$, whose action on the states 
$\big | j \big \rangle$
is 
$\hat{X} \big | j \big \rangle \;=\; \big | j+1 \, (mod \, d) \big \rangle$
and 
$\hat{Z} \big | j \big \rangle \;=\; \Omega^j \, \big | j \big \rangle$.
Here $\Omega \;=\; e^{\frac{2\pi i}{d}}$.  
The extension of the qubit Bloch representation for a density matrix $\rho$ 
to qudits is shown in appendix A to be
$$
\rho \;=\; \frac{1}{d} \, \sum_{a,b\in 
\{0,1,2,\cdots,d-1\}} \; \alpha_{a,b} \; \hat{X}^a \; \hat{Z}^b.
$$ 
The $\alpha_{a,b}$ are complex quantities. Define
$E_{a,b} \,=\, \hat{X}^a \hat{Z}^b$.
Note that $E_{0,0} \,=\, I_d$. 
In appendix A it is shown
$Trace(E_{a,b}) \;=\; d \, \delta_{a,0} \, \delta_{b,0}$, where $\delta$
is the Kronecker delta function.
The trace condition $Trace(\rho) \;=\; 1$ allows us to conclude
$\alpha_{0,0} \;=\; 1$. 
Let $\Upsilon$ denote the set of $d^2 \,-\,1$ elements 
$a,b \;\in\; \{0,1,2,\cdots,d-1\}$  {\it with the exception that 
$a$ and $b$ cannot both be zero.} Then we can write the qudit 
density matrix $\rho$ as 
$\rho \;=\; \frac{1}{d} \, \left ( I_d \;+\; \sum_{(a,b) \in \Upsilon } \; 
\alpha_{a,b} \; E_{a,b} \right ) $.  
A qudit quantum channel $\mathcal{E}$ is a linear map.
One can write such a map as a $d^2$ by $d^2$ complex 
matrix $\mathcal{M}$ taking the $d^2$ vector of 
coefficients 
$\alpha_{a,b}$ of $\rho$ to the $d^2$ set of
coefficients $\widetilde{\alpha}_{a,b}$ of 
$\widetilde{\rho}\;=\; \mathcal{E}(\rho)$. 
\footnote{Our qudit matrix development in which we 
write $\mathcal{E}$ as a $d^2$ by $d^2$ 
matrix closely follows work done in \cite{rsw} 
for the unital qubit channel case.}

If the qudit quantum channel $\mathcal{E}$ is 
unital, meaning $\mathcal{E}(I_d)\;=\; I_d$, then 
the first row and column of $\mathcal{M}$ are
a one followed by $d^2-1$ zeros. Hence we can represent a 
qudit unital channel by a matrix $\mathcal{N}$ of
$d^2\,-\,1$ 
by 
$d^2\,-\,1$ 
complex entries mapping the vector of $d^2\,-\,1$ coefficients 
$\alpha_{(a,b)}$, with $(a,b)\in\Upsilon$, representing $\rho$ to
the vector of $d^2\,-\,1$ coefficients 
$\widetilde{\alpha}_{(a,b)}$, with $(a,b)\in\Upsilon$,
representing $\widetilde{\rho}\;=\; \mathcal{E}(\rho)$.  
The specific class of qudit channels we shall be interested in 
are those completely positive unital quantum channels 
for which $\mathcal{N}$ is diagonal. This class of channels is 
nonempty. For example, consider the channel corresponding to
all zeros on the diagonal.
This point channel maps all input density 
matrices to a single output density matrix
$\widetilde{\rho}\;=\; \frac{1}{d} \, I_d$. 
Another member of the set of diagonal unital channels is the identity 
map, which maps any input density matrix to itself. This channel 
has all ones on the diagonal of the matrix $\mathcal{N}$.
 
The approach we take to determine
the HSW channel capacity for this special class of diagonal 
unital qudit channels closely follows our unital qubit channel 
derivation above.
Note the operators $E_{a,b}$ are unitary.
Using the commutation relation shown in appendix A, 
$\hat{Z}\hat{X} \,=\, \Omega \hat{Z}\hat{X}$, where
$\Omega \,=\, e^{\frac{2 \pi i }{d}}$,
we have
\begin{equation}\label{E:8}
E_{g,h} \, E_{a,b} \, E_{g,h}^{\dagger} \;=\;
\hat{X}^g \hat{Z}^h \, \hat{X}^a \hat{Z}^b \, 
\hat{Z}^{-h} \hat{X}^{-g}
\,=\, 
\Omega^{ah} \, \hat{X}^g \hat{X}^a \, \hat{Z}^h \hat{Z}^b \, 
\hat{Z}^{-h} \hat{X}^{-g}
\end{equation}
$$
\,=\, 
\Omega^{ah} \, \hat{X}^g \hat{X}^a \hat{Z}^b \hat{X}^{-g}
\,=\,
\Omega^{ah} \, \Omega^{-bg} \, \hat{X}^g \hat{X}^a \hat{X}^{-g} \hat{Z}^b 
\,=\, 
\Omega^{ah} \, \Omega^{-bg} \, \hat{X}^a \hat{X}^g \hat{X}^{-g} \hat{Z}^b 
$$
$$
\,=\, 
\Omega^{ah} \, \Omega^{-bg} \, \hat{X}^a \hat{Z}^b
\,=\, 
\Omega^{ah \,-\, bg} \, E_{a,b}.
$$ 
Define $F_{a,b,c} \;=\; \Omega^c \, E_{a,b}$, where
$a,b,c \in \big \{ 0,1,2,\cdots,d-1\big \}$. 
Since $\Omega^c$ and the $E_{a,b}$ are unitary operators,  
$F_{a,b,c}$ is a unitary operator. 
The action of the $F_{a,b,c}$ on a diagonal unital qudit channel output 
density operator
$\widetilde{\rho}$ is 
\begin{equation}\label{E:12}
F_{a,b,c} \; \widetilde{\rho} \; F_{a,b,c}^{\dagger} \;=\; 
E_{a,b} \; \widetilde{\rho} \;  E_{a,b}^{\dagger} \;=\; 
E_{a,b} \; \mathcal{E}(\rho) \; E_{a,b}^{\dagger}
\end{equation}
$$
\;=\; 
E_{a,b} \; \frac{1}{d} \, \left ( I_d \;+\; \sum_{(q,r) \in \Upsilon } \; 
\lambda_{q,r} \, \alpha_{q,r} \; E_{q,r} \right ) \;  E_{a,b}^{\dagger} \;=\; 
\frac{1}{d} \, \left ( I_d \;+\; \sum_{(q,r) \in \Upsilon } \; 
\lambda_{q,r} \, \alpha_{q,r} \; 
E_{a,b} \,E_{q,r} \,  E_{a,b}^{\dagger} \right ) 
$$
$$
\;=\; 
\frac{1}{d} \, \left ( I_d \;+\; \sum_{(q,r) \in \Upsilon } \; 
\lambda_{q,r} \, \alpha_{q,r} \; \Omega^{bq-ar} \, E_{q,r} \right ) \;=\; 
\mathcal{E}\left ( \frac{1}{d} \, \left ( I_d \;+\; \sum_{(q,r) \in \Upsilon } \; 
\alpha_{q,r} \; \Omega^{bq-ar} \, E_{q,r} \right ) \right ) 
$$
$$\;=\; 
\mathcal{E}\left ( 
E_{a,b} \; \frac{1}{d} \, \left ( I_d \;+\; \sum_{(q,r) \in \Upsilon } \; 
\alpha_{q,r} \; E_{q,r} \right ) E_{a,b}^{\dagger} \right ) \;=\; 
\mathcal{E}\left ( E_{a,b} \; \rho \; E_{a,b}^{\dagger} \right ) \;=\; 
\mathcal{E}\left ( F_{a,b} \; \rho \; F_{a,b,c}^{\dagger} \right ) .
$$
Since the $F_{a,b,c}$ are unitary operators, we conclude that given 
any optimal input ensemble \oe, the output ensemble
$\Theta_{a,b,c}$ obtained by applying $F_{a,b,c}$ to \oo is achievable
and $\Theta_{a,b,c}$ has the optimal input ensemble 
$\Big \{p_i \,,\, \phi_i \,=\, F_{a,b,c} \, \rho_i \, F_{a,b,c}^{\dagger} \Big \}$.
Each of the $\phi_i$ is a valid input density operator due to the fact that
$F_{a,b,c}$ is a unitary operator, and is implementing
a change of basis on 
$\rho_i$. 

The set of operators $\{ F_{a,b,c}\}$ forms a group of order $d^3$ 
which we shall call $\mathcal{Q}$.  
Recall our theorem for proving a finite group is reducible.\cite{cornwell}
The group $\mathcal{Q}$
is reducible since $\Big | \, Trace\Big [ F_{a,b,c}\Big ]\, \Big |$
equals zero when either $a$ and $b$ are non-zero, and
$\Big | \, Trace\Big [ F_{a,b,c}\Big ]\, \Big |$ equals $d$ when $a=b=0$.
Thus
$\frac{1}{\| \mathcal{Q}\|} 
\sum_{q\in\mathcal{Q}} \, \Big \vert Trace[q] \Big | ^2 \,=\, 
\frac{1}{d^3}  \, d \, d^2 \,=\, 1$.  
Since $\mathcal{Q}$ is a reducible group, we can apply Schur's lemma.
For any optimal input ensemble \oe, 
the channel output ensemble $\Big \{ p_i \,,\, F_{a,b,c} \,
\mathcal{E}(\rho_i) \,  F_{a,b,c}^{\dagger} \Big \}$ is achievable and 
the corresponding input ensemble 
$\Big \{ p_i \,,\, F_{a,b,c} \, \rho_i \, F_{a,b,c}^{\dagger}\Big \}$ is optimal. 
From the uniqueness of the average output state 
$\widetilde{\Phi}$ for any optimal ensemble,
we conclude that 
$\forall \; a,b,c\;:\; F_{a,b,c} \, \widetilde{\Phi} \, F_{a,b,c}^{\dagger} \,=\,  
\widetilde{\Phi}$ or
$F_{a,b,c} \, \widetilde{\Phi} \, = \,  
\widetilde{\Phi} \, F_{a,b,c}$. By Schur's lemma we obtain 
$\widetilde{\Phi} \, \propto \, I_d$. The trace condition
tells us $Trace\left ( \widetilde{\Phi} \right ) \,=\, 1$, so 
we conclude $\widetilde{\Phi} \,\equiv\, \frac{1}{d} \, I_d$. 

This leads us to conclude that for the optimal input ensemble \oe, the HSW 
channel capacity is 
$\mathcal{C} \;=\; \log_2(d) \;-\; 
\sum_i \; p_i \mathcal{S}(\mathcal{E}(\rho_i))$. 
Using the two relative entropy results from equations 
\ref{E:4},\ref{E:5},\ref{E:6}, and \ref{E:7},
as we did in the qubit case, we obtain 
$
\mathcal{S}(\mathcal{E}(\rho_i))\; =\;  
min_{\phi} \, \mathcal{S}\left ( \mathcal{E}(\phi) \right ) 
$
yielding the HSW channel capacity for
diagonal unital qudit channels. 
$$
\mathcal{C} \;=\; \log_2(d) \;-\; 
min_{\phi} \, \mathcal{S}\left ( \mathcal{E}(\phi) \right ) .
$$

\section{Products of Diagonal Unital Qudit Channels} 

Consider the product of N diagonal unital qudit channels 
$\mathcal{E}^{(k)}\, , \, k=1,\cdots ,N$. 
The tensor product channel is
$
\mathcal{E}^{\otimes} \,=\, 
\mathcal{E}^{(1)} \otimes 
\mathcal{E}^{(2)} \otimes 
\cdots 
\otimes \mathcal{E}^{(N)} 
$.
Let the input qudit density operator $\rho^{(k)}$ 
corresponding to the diagonal unital channel $\mathcal{E}^{(k)}$   
be of dimension $d_k$. 
Then $d\,=\, \prod_{k=1}^{N} \,d_k$ is the dimension of the 
input qudit $\rho^{\otimes}$ for the product channel
$\mathcal{E}^{\otimes}$. 
The basis elements for $\rho^{\otimes}$
which we shall use are the tensor products
of the individual $E_{a,b}^{(k)}$.  
$$
\left \{ E_{a,b}^{\otimes } \right \} \,=\, \left \{
E_{a_1,b_1}^{(1)} \otimes
E_{a_2,b_2}^{(2)} \otimes 
\cdots \otimes 
E_{a_N,b_N}^{(N)}  \right \} ,
$$ 
where the $a_k$ and $b_k \,\in \{0,1,2,\ldots,d_k-1\}$ and
$a$ and $b \,\in \{0,1,2,\ldots,d-1\}$.

The basis elements $E_{a,b}^{\otimes} $ are not necessarily
constructed using the $d$ dimensional qudit operators 
$\hat{X}$ and $\hat{Z}$ described in appendix A. As a result, we must 
prove several properties for the set $\Big \{ E_{a,b}^{\otimes}\Big \}$ before 
we proceed with the HSW channel capacity analysis for product channels. 

\subsection{The mapping $\{a_1,a_2,a_3,\cdots,a_{N-1},a_{N}\} \Longleftrightarrow \{a\}$.}

The set of possible coefficients 
$\{a_1,a_2,a_3,\cdots,a_{N-1},a_{N}\}$
and the set of possible $\{a^{\otimes}\}$
both have $d$ elements, where 
$d=\prod_{k=1}^{k=N} d_k$. Here the $\{a_k\} \in 
\big \{ 0,1,2,\cdots,d_k-1\big \}$ and
$\{a\} \in \big \{ 0,1,2,\cdots,d \big \}$.
There are many bijective mappings between these two sets, and
it is useful to have one particular map in mind as we proceed.
The one we shall use is presented in the table below.
\begin{center}
\begin{tabular}{||c||c|c|c|c|c||c|c|c|c|c||c|c|c||}
\hline
\hline
$a_N$ & 0 & 0 & 0 & $\cdots$ & 0 & 0 & 0 & 0 & $\cdots$ & 0 & 0 & 
0 & $\cdots$ \\ 
\hline
$a_{N-1}$ & 0 & 0 & 0 & $\cdots$ & 0 & 0 & 0 & 0 & $\cdots$ & 0 & 
0 & 0 & $\cdots$ \\ 
\hline
$\vdots$ & $\vdots$ & $\vdots$ & $\vdots$ & $\vdots$ & $\vdots$ & 
$\vdots$ & $\vdots$ & $\vdots$ & $\vdots$ & $\vdots$ & $\vdots$ & 
$\vdots$ & $\cdots$ \\
\hline
$a_3$ & 0 & 0 & 0 & $\cdots$ & 0 &   0 & 0 & 0 & $\cdots$ & 0 & 
0 & 0 & $\cdots$ \\ 
\hline
$a_2$ & 0 & 0 & 0 &  $\cdots$ & 0 & 1 & 1 & 1 &  $\cdots$ & 1 & 2 & 
2 & $\cdots$ \\
\hline
$a_1$ & 0 & 1 & 2  & $\cdots$ & $d_1-1$ 
 & 0 & 1 & 2  & $\cdots$ & $d_1-1$ & 0 & 1 & $\cdots$ \\
\hline
\hline
$a$ & 0 & 1 & 2 & $\cdots$ & $d_1-1$ & $d_1$ &$d_1$+1  &$d_1$+2 
& $\cdots$ & $2d_1-1$ & $2d_1$ & $2d_1+1$ &  $\cdots$ \\
\hline
\hline
\end{tabular}
\end{center}
and so on. Below, we associate an $E_{a,b}^{\otimes}$ with the
tensor product $\left \{
E_{a_1,b_1}^{(1)} \otimes
E_{a_2,b_2}^{(2)} \otimes 
\cdots \otimes 
E_{a_N,b_N}^{(N)}  \right \} 
$ by using this mapping twice, once for the association
$\{a_1,a_2,a_3,\cdots,a_{N-1},a_{N}\}\Longleftrightarrow \{a^{\otimes}\}$
amd again for
$\{b_1,b_2,b_3,\cdots,b_{N-1},b_{N}\}\Longleftrightarrow \{b^{\otimes}\}$.

\subsection{Orthonormality of the $\{E_{a,b}^{\otimes}\}$.} 

The operators $E_{a,b}^{\otimes}$ form, with respect to the
Hilbert-Schmidt norm, a set of $d^2$ orthogonal
operators. The orthogonality
of the $\Big \{ E_{a,b}^{\otimes}\Big \}$
is inherited from 
the orthogonality of the operators $\Big \{E_{a_k,b_k}^{(k)}\Big \}$,
which is shown in appendix A, equation (\ref{E:15}). Using properties of 
tensors from \cite{halmosII}, we have  
\begin{equation}\label{E:9}
\Big \langle E_{a,b}^{\otimes}, E_{g,h}^{\otimes} \Big \rangle \,=\, 
Trace \Big [ E_{a,b}^{\otimes^{\dagger}}\, E_{g,h}^{\otimes} \Big ]  
\end{equation}
$$
=\, Trace \left [ 
\left ( 
E_{a_1,b_1}^{(1)} \otimes
E_{a_2,b_2}^{(2)} \otimes
\cdots \otimes
E_{a_N,b_N}^{(N)} \right )^{\dagger} \left ( 
E_{g_1,h_1}^{(1)} \otimes 
E_{g_2,h_2}^{(2)} \otimes 
\cdots \otimes
E_{g_N,h_N}^{(N)}  \right ) \right ] 
$$ 
$$
= \, Trace \left [ 
\left ( 
E_{a_1,b_1}^{(1)^{\dagger}} \otimes
E_{a_2,b_2}^{(2)^{\dagger}} \otimes
\cdots \otimes
E_{a_N,b_N}^{(N)^{\dagger}} \right ) \left ( 
E_{g_1,h_1}^{(1)} \otimes 
E_{g_2,h_2}^{(2)} \otimes 
\cdots \otimes
E_{g_N,h_N}^{(N)}  \right ) \right ] 
$$ 
$$
= \, Trace \left [ 
\left ( 
E_{a_1,b_1}^{(1)^{\dagger}} 
E_{g_1,h_1}^{(1)} 
\right ) \otimes 
\left ( 
E_{a_2,b_2}^{(2)^{\dagger}} 
E_{g_2,h_2}^{(2)} 
\right ) \otimes \cdots \otimes  
\left ( 
E_{a_N,b_N}^{(N)^{\dagger}} 
E_{g_N,h_N}^{(N)} 
\right ) 
\right ] 
$$ 
$$
=\, Trace \left [  
E_{a_1,b_1}^{(1)^{\dagger}} 
E_{g_1,h_1}^{(1)} 
\right ] \, 
Trace \left [ 
E_{a_2,b_2}^{(2)^{\dagger}} 
E_{g_2,h_2}^{(2)} 
\right ] \, \cdots \,  Trace \left [ 
E_{a_N,b_N}^{(N)^{\dagger}} 
E_{g_N,h_N}^{(N)} 
\right ] 
$$ 
$$
= \,
\left ( d_1 \, 
\delta_{a_1,g_1} 
\delta_{b_1,h_1} 
\right ) 
\, 
\left (  d_2 \, 
\delta_{a_2,g_2} 
\delta_{b_2,h_2} 
\right ) 
\, \cdots \, 
\left (  d_N \, 
\delta_{a_N,g_N} 
\delta_{b_N,h_N} 
\right )
\,=\, d \, \delta_{a,g} \, \delta_{b,h},
$$ 
where we used our map between the sets 
$\{a^{(k)},b^{(k)}\} \rightarrow \{a^{\otimes},b^{\otimes}\}$,
and the fact $d=\prod_{k=1}^{k=N} d_k$.
Thus we conclude 
$ \Big \langle E_{a,b}^{\otimes}, E_{g,h}^{\otimes} \Big \rangle \,=\, 
Trace \Big [ E_{a,b}^{\otimes\dagger}, E_{g,h}^{\otimes} \Big ]  
\,=\,  \delta_{a,g} \delta_{b,h}$. 
The orthogonality of the $\Big \{ E_{a,b}^{\otimes}\Big \} $ means we can 
expand $\rho^{\otimes}$ in terms of the $\Big \{ E_{a,b}^{\otimes}\Big \}$, 
yielding
$\rho^{\otimes} \;=\; \frac{1}{d} \, \sum_{a,b\in 
\{0,1,2,\cdots,d-1\}} \; \alpha_{a,b} \; E_{a,b}^{\otimes}
$.

Another property of the $E_{a,b}^{\otimes}$ we shall need is the result of
$E_{g,h}^{\otimes} \, E_{a,b}^{\otimes} \, E_{g,h}^{\otimes^\dagger}$.
Using equation (\ref{E:8}), and the tensor nature of
$E_{a,b}^{\otimes}$, we have
$E_{g,h}^{\otimes} \, E_{a,b}^{\otimes} \, E_{g,h}^{\otimes^\dagger}\;=\;$
$$
\left ( 
E_{g_1,h_1}^{(1)} \otimes 
E_{g_2,h_2}^{(2)} \otimes 
\cdots \otimes
E_{g_N,h_N}^{(N)}  \right )
\left ( 
E_{a_1,b_1}^{(1)} \otimes
E_{a_2,b_2}^{(2)} \otimes
\cdots \otimes
E_{a_N,b_N}^{(N)} \right )
\left ( 
E_{g_1,h_1}^{(1)} \otimes 
E_{g_2,h_2}^{(2)} \otimes 
\cdots \otimes
E_{g_N,h_N}^{(N)}  \right )^{\dagger}
$$
$$
\;=\;
\left ( 
E_{g_1,h_1}^{(1)} \otimes 
E_{g_2,h_2}^{(2)} \otimes 
\cdots \otimes
E_{g_N,h_N}^{(N)}  \right )
\left ( 
E_{a_1,b_1}^{(1)} \otimes
E_{a_2,b_2}^{(2)} \otimes
\cdots \otimes
E_{a_N,b_N}^{(N)} \right )
\left ( 
E_{g_1,h_1}^{(1)^\dagger} \otimes 
E_{g_2,h_2}^{(2)^\dagger} \otimes 
\cdots \otimes
E_{g_N,h_N}^{(N)^\dagger}  \right )
$$
$$
\;=\;
\left ( 
E_{g_1,h_1}^{(1)} 
E_{a_1,b_1}^{(1)} 
E_{g_1,h_1}^{(1)^{\dagger}} 
\right ) \otimes 
\left ( 
E_{g_2,h_2}^{(2)} 
E_{a_2,b_2}^{(2)} 
E_{g_2,h_2}^{(2)^{\dagger}} 
\right ) \otimes 
\cdots \otimes
\left ( 
E_{g_N,h_N}^{(N)} 
E_{a_N,b_N}^{(N)} 
E_{g_N,h_N}^{(N)^{\dagger}} 
\right )
$$
$$
\;=\;
\left ( 
{\omega_1}^{a_1 h_1 - b_1 g_1} \, 
E_{a_1,b_1}^{(1)} 
\right ) \otimes 
\left ( 
{\omega_2}^{a_2 h_2 - b_2 g_2} \, 
E_{a_2,b_2}^{(2)} 
\right ) \otimes 
\cdots \otimes
\left ( 
{\omega_N}^{a_N h_N - b_N g_N} \, 
E_{a_N,b_N}^{(N)} 
\right )
$$
\begin{equation}\label{E:10}
\;=\;
\Omega^c \, 
E_{a_1,b_1}^{(1)} 
\otimes 
E_{a_2,b_2}^{(2)} 
\otimes 
\cdots \otimes
E_{a_N,b_N}^{(N)} 
\;=\;
\Omega^c \, 
E_{a,b}^{\otimes},
\end{equation}
where $\omega_k \,=\, e^{\frac{2 \pi i}{d_k}}$,
$\Omega \,=\, e^{\frac{2 \pi i}{d}}$, and
$c \,=\, \sum_{k=1}^{k=N} \; ( a_k h_k - b_k g_k )  \frac{d}{d_k}$.  

\subsection{
The channel $\mathcal{E}^{\otimes}$ is unital and diagonal 
in the $E_{a,b}^{\otimes}$ basis.}

The channel $\mathcal{E}^{\otimes}$ is diagonal 
in the $E_{a,b}^{\otimes}$ basis. To see this, note that 
\begin{equation}\label{E:11}
\mathcal{E}^{\otimes} \left ( E_{a,b}^{\otimes} \right ) \,=\, 
\mathcal{E}^{\otimes} \left (
E_{a_1,b_1}^{(1)} \otimes
E_{a_2,b_2}^{(2)} \otimes
\cdots \otimes
E_{a_N,b_N}^{(N)} \right ) 
\end{equation}
$$
\,=\, 
\mathcal{E}^{(1)} \otimes 
\mathcal{E}^{(2)} \otimes 
\cdots 
\otimes \mathcal{E}^{(N)} 
\left ( 
E_{a_1,b_1}^{(1)} \otimes
E_{a_2,b_2}^{(2)} \otimes
\cdots \otimes
E_{a_N,b_N}^{(N)} \right ) 
$$
$$
\,=\, 
\mathcal{E}^{(1)}\left ( E_{a_1,b_1}^{(1)} \right )  \otimes 
\mathcal{E}^{(2)}\left ( E_{a_2,b_2}^{(2)} \right )  \otimes \cdots \otimes 
\mathcal{E}^{(N)}\left ( E_{a_N,b_N}^{(N)} \right )  
$$
$$
\,=\, 
\left ( \alpha_{a_1,b_1}^{(1)} \, E_{a_1,b_1}^{(1)} \right )  \otimes 
\left ( \alpha_{a_2,b_2}^{(2)} \, E_{a_2,b_2}^{(2)} \right )  \otimes 
\cdots \otimes 
\left ( \alpha_{a_N,b_N}^{(N)} \, E_{a_N,b_N}^{(N)} \right )  
$$
$$
\,=\, 
\alpha_{a_1,b_1}^{(1)} \, \alpha_{a_2,b_2}^{(2)} \, 
\cdots \, \alpha_{a_N,b_N}^{(N)} \, 
\left ( E_{a_1,b_1}^{(1)} \right )  \otimes 
\left ( E_{a_2,b_2}^{(2)} \right )  \otimes 
\cdots \otimes 
\left ( E_{a_N,b_N}^{(N)} \right )  
$$
$$
\,=\, 
\Lambda_{a,b} \, 
\left ( E_{a_1,b_1}^{(1)} \right )  \otimes 
\left ( E_{a_2,b_2}^{(2)} \right )  \otimes 
\cdots \otimes 
\left ( E_{a_N,b_N}^{(N)} \right )  
\,=\, 
\Lambda_{a,b} \, 
E_{a,b}^{\otimes},
$$ 
where
$
\Lambda_{a,b} \,=\,  
\alpha_{a_1,b_1}^{(1)} \, \alpha_{a_2,b_2}^{(2)} \, 
\cdots \, \alpha_{a_N,b_N}^{(N)}$, and we used our bijective map
$\Big \{a^{(k)},b^{(k)}\Big \} \Longleftrightarrow \{a^{\otimes},b^{\otimes}\}$
to move back and forth between the operator basis set
$\Big \{ E_{a,b}^{\otimes}\Big \}$
and the operator basis set 
$\Big \{ E_{a_1,b_1}^{(1)}  \otimes 
E_{a_2,b_2}^{(2)}  \otimes 
\cdots \otimes E_{a_N,b_N}^{(N)}\Big \}$.
Thus the tensor product of diagonal qudit channels yields a diagonal 
qudit channel.

Next note that 
$
E_{0,0}^{\otimes} \,=\, 
E_{0,0}^{(1)} \otimes 
E_{0,0}^{(2)} \otimes \cdots \otimes 
E_{0,0}^{(N)} \,=\, 
I_{d_1} \otimes I_{d_2} \otimes \cdots \otimes I_{d_N}\, =\,  I_d.
$
Taking a special case of the result in equation (\ref{E:11}), we obtain 
$$
\mathcal{E}^{\otimes} \left ( I_d  \right ) \,=\, 
\mathcal{E}^{\otimes} \left ( E_{0,0}^{\otimes} \right ) \,=\, 
\mathcal{E}^{(1)} \left ( E_{0,0}^{(1)} \right )  \otimes
\mathcal{E}^{(2)} \left ( E_{0,0}^{(2)} \right )  \otimes \cdots \otimes
\mathcal{E}^{(N)} \left ( E_{0,0}^{(N)} \right )  
$$
$$
\,=\, 
\mathcal{E}^{(1)} \left ( I_{d_1} \right )  \otimes
\mathcal{E}^{(2)} \left ( I_{d_2} \right )  \otimes \cdots \otimes 
\mathcal{E}^{(N)} \left ( I_{d_N} \right )
\, =\, 
I_{d_1} \otimes I_{d_2} \otimes \cdots \otimes I_{d_N}\,=\, I_d.
$$ 
We conclude that
$\mathcal{E}^{\otimes} \left ( I_d \right )  
\,=\, I_d$, and the channel $\mathcal{E}^{\otimes}$ is unital.
Thus
the tensor product of diagonal, unital qudit channels yields a diagonal
unital qudit channel.

As an example, consider the product of two qubit (diagonal) unital channels,
$\mathcal{E}^{(1)}$ with diagonal parameters 
$\{\lambda_1 \,,\, \lambda_2 \,,\, \lambda_3 \}$, and
$\mathcal{E}^{(2)}$ with diagonal parameters 
$\{\xi_1 \,,\, \xi_2 \,,\, \xi_3 \}$. The product channel
$\mathcal{E}^{\otimes} \,=\, \mathcal{E}^{(1)} \otimes \mathcal{E}^{(2)}$ 
is a diagonal, unital channel, taking an input vector of
$(d_1 \, d_2)^2 -1 \,=\, 4^2-1 \,=\, 15$ 
input density matrix
coefficients $\alpha_{a,b}$ to the 
output density matrix
coefficients $\widetilde{\alpha}_{a,b}$, as shown below.  
$$
\cmatrix{
\{ \;basis \;  element\; I_2 \otimes \sigma_x \;\}  \cr  
\{ \;basis \;  element\; I_2 \otimes \sigma_y \;\}  \cr  
\{ \;basis \;  element\; I_2 \otimes \sigma_z \;\}  \cr
\{ \;basis \;  element\; \sigma_x \otimes I_2 \;\}  \cr  
\{ \;basis \;  element\; \sigma_y \otimes I_2 \;\}  \cr  
\{ \;basis \;  element\; \sigma_z \otimes I_2 \;\}  \cr
\{ \;basis \;  element\; \sigma_x \otimes \sigma_x \;\}  \cr  
\{ \;basis \;  element\; \sigma_x \otimes \sigma_y \;\}  \cr  
\{ \;basis \;  element\; \sigma_x \otimes \sigma_z \;\}  \cr
\{ \;basis \;  element\; \sigma_y \otimes \sigma_x \;\}  \cr  
\{ \;basis \;  element\; \sigma_y \otimes \sigma_y \;\}  \cr  
\{ \;basis \;  element\; \sigma_y \otimes \sigma_z \;\}  \cr
\{ \;basis \;  element\; \sigma_z \otimes \sigma_x \;\}  \cr  
\{ \;basis \;  element\; \sigma_z \otimes \sigma_y \;\}  \cr  
\{ \;basis \;  element\; \sigma_z \otimes \sigma_z \;\}  
}
\qquad 
\bmatrix{ 
\alpha_{0,1} \cr
\alpha_{0,2} \cr
\alpha_{0,3} \cr 
\alpha_{1,0} \cr
\alpha_{2,0} \cr
\alpha_{3,0} \cr 
\alpha_{1,1} \cr
\alpha_{1,2} \cr
\alpha_{1,3} \cr 
\alpha_{2,1} \cr
\alpha_{2,2} \cr
\alpha_{2,3} \cr
\alpha_{3,1} \cr
\alpha_{3,2} \cr
\alpha_{3,3} 
}
\quad   \stackrel{\mathcal{E}}{\longrightarrow} \quad
\bmatrix{
\widetilde{\alpha}_{0,1} \,=\, \phantom{\lambda_1} \, \xi_1 \, \alpha_{0,1} \cr
\widetilde{\alpha}_{0,2} \,=\, \phantom{\lambda_1} \, \xi_2 \, \alpha_{0,2} \cr
\widetilde{\alpha}_{0,3} \,=\, \phantom{\lambda_1} \, \xi_3 \, \alpha_{0,3} \cr
\widetilde{\alpha}_{1,0} \,=\, \lambda_1 \, \phantom{\xi_1} \, \alpha_{1,0} \cr
\widetilde{\alpha}_{2,0} \,=\, \lambda_2 \, \phantom{\xi_1} \, \alpha_{2,0} \cr
\widetilde{\alpha}_{3,0} \,=\, \lambda_3 \, \phantom{\xi_1} \, \alpha_{3,0} \cr
\widetilde{\alpha}_{1,1} \,=\, \lambda_1 \, \xi_1 \, \alpha_{1,1} \cr
\widetilde{\alpha}_{1,2} \,=\, \lambda_1 \, \xi_2 \, \alpha_{1,2} \cr
\widetilde{\alpha}_{1,3} \,=\, \lambda_1 \, \xi_3 \, \alpha_{1,3} \cr
\widetilde{\alpha}_{2,1} \,=\, \lambda_2 \, \xi_1 \, \alpha_{2,1} \cr
\widetilde{\alpha}_{2,2} \,=\, \lambda_2 \, \xi_2 \, \alpha_{2,2} \cr
\widetilde{\alpha}_{2,3} \,=\, \lambda_2 \, \xi_3 \, \alpha_{2,3} \cr
\widetilde{\alpha}_{3,1} \,=\, \lambda_3 \, \xi_1 \, \alpha_{3,1} \cr
\widetilde{\alpha}_{3,2} \,=\, \lambda_3 \, \xi_2 \, \alpha_{3,2} \cr
\widetilde{\alpha}_{3,3} \,=\, \lambda_3 \, \xi_3 \, \alpha_{3,3} 
}
$$

\subsection{The average output state of an optimal ensemble 
$\widetilde{\Phi}$ is $\propto I_d$ for $\mathcal{E}^{\otimes}$.}

Define the set of $d^3$ operators 
$\big \{F_{a,b,c}^{\otimes}\big \}$ as 
$
F_{a,b,c}^{\otimes} \,=\, 
e^{\frac{2 \pi i c  }{d}}  \, E_{a,b}^{\otimes}$.
Using our bijective map between 
$\{a_1,a_2,a_3,\cdots,a_{N-1},a_{N}\}$ and $\{a\}$, 
we expand $F_{a,b,c}^{\otimes}$ in terms of a phase $e^{\frac{2 \pi i}{d}}$ and
the $\Big \{E_{a_k,b_k}^{(k)}\Big \}$.
Our expression for 
$F_{a,b,c}^{\otimes}$ becomes
$$
F_{a,b,c}^{\otimes} \,=\, 
e^{\frac{2 \pi i c  }{d}}  \, E_{a,b}^{\otimes}
\,=\,
e^{\frac{2 \pi i c  }{d}}  \,
E_{a_1,b_1}^{(1)} \otimes
E_{a_2,b_2}^{(2)} \otimes \cdots \otimes 
E_{a_N,b_N}^{(N)}.
$$ 
The set of operators $\big \{F_{a,b,c}^{\otimes}\big \}$ 
are the product of a phase 
$e^{\frac{2 \pi i c  }{d}}$
and the tensor products of the 
individual operators $\Big \{E_{a_k,b_k}^{(k)}\Big \}$.
The $\big \{F_{a,b,c}^{\otimes}\big \}$ are unitary operators, inheriting this behavior 
from the unitary nature of the phase factor and the 
unitary nature of the subsystem operators $\Big \{E_{a_k,b_k}^{(k)}\Big \}$. 
To see this, note 
$$
F_{a,b,c}^{\otimes^{\dagger}}\, F_{a,b,c}^{\otimes}
=\, 
\left ( 
e^{\frac{2 \pi i c  }{d}}  \,
E_{a_1,b_1}^{(1)} \otimes
E_{a_2,b_2}^{(2)} \otimes
\cdots \otimes
E_{a_N,b_N}^{(N)} \right )^{\dagger} \left ( e^{\frac{2 \pi i c  }{d}}  \,
E_{a_1,b_1}^{(1)} \otimes 
E_{a_2,b_2}^{(2)} \otimes 
\cdots \otimes
E_{a_N,b_N}^{(N)}  \right ) 
$$ 
$$
= \,  
e^{\frac{-2 \pi i c  }{d}}  \,
e^{\frac{2 \pi i c  }{d}}  \,
\left ( 
E_{a_1,b_1}^{(1)^{\dagger}} \otimes
E_{a_2,b_2}^{(2)^{\dagger}} \otimes
\cdots \otimes
E_{a_N,b_N}^{(N)^{\dagger}} \right ) \left ( 
E_{a_1,b_1}^{(1)} \otimes 
E_{a_2,b_2}^{(2)} \otimes 
\cdots \otimes
E_{a_N,b_N}^{(N)}  \right ) 
$$ 
$$
= \,
1 \, 
\left ( 
E_{a_1,b_1}^{(1)^{\dagger}} 
E_{a_1,b_1}^{(1)} 
\right ) \otimes 
\left ( 
E_{a_2,b_2}^{(2)^{\dagger}} 
E_{a_2,b_2}^{(2)} 
\right ) \otimes \cdots \otimes  
\left ( 
E_{a_N,b_N}^{(N)^{\dagger}} 
E_{a_N,b_N}^{(N)} 
\right )  
$$
$$
\,=\, 
I_{d_1} \otimes I_{d_2} \otimes \cdots \otimes I_{d_N} \,=\,
I_d,$$
where we used the unitary nature of the 
individual $\Big \{E_{a_k,b_k}^{(k)}\Big \}$
to say $E_{a_k,b_k}^{(k)^{\dagger}}\, E_{a_k,b_k}^{(k)}\,=\, 
I_{d_k}$.

The $\big \{F_{a,b,c}^{\otimes}\big \}$ form an 
irreducible group which we shall call $\mathcal{Q}$.
To see why $\mathcal{Q}$ is irreducible,
recall our relation for irreducibility 
from \cite{cornwell} discussed above. 
A necessary and sufficient condition for a finite group
$\mathcal{G}$ to be irreducible is if
the relation  $\frac{1}{\| \mathcal{G}\|} 
\sum_{g\in\mathcal{G}} \, \Big \vert Trace[g] \Big | ^2 \;=\; 1$
is true\cite{cornwell}. 
Here $\|\mathcal{G}\|$ is the order of the group $\mathcal{G}$.
Let the group $\mathcal{Q}$ be the 
set $\big \{F_{a,b,c}^{\otimes}\big \}$, where
$a,b,c \in \{ 0,1,2,\ldots,d-1\}$. $\mathcal{Q}$ is of order $d^3$ and 
hence finite. Previously, we noted that 
$E_{0,0}^{\otimes} \,=\, I_d$ and 
$Trace \Big [ E_{a,b}^{\otimes\dagger}, E_{g,h}^{\otimes} \Big ]  
\,=\,  \delta_{a,g} \delta_{b,h}$. Thus
$Trace \left [ E_{a,b}^{\otimes} \right ] \,=\, 
d \, \delta_{a,0} \, \delta_{b,0}$. 
Computing the Trace sum yields 
$$
\frac{1}{\| \mathcal{Q}\|} 
\sum_{q\in\mathcal{Q}} \, \Big \vert Trace[q] \Big | ^2 \,=\,
\frac{1}{d^3} 
\sum_{c \in\{0,1,2,\ldots,d-1\} } \, 
\sum_{b \in\{0,1,2,\ldots,d-1\} } \, 
\sum_{a \in\{0,1,2,\ldots,d-1\} } \, 
\Big \vert Trace[F_{a,b,c}^{\otimes} ] \Big | ^2 
$$
$$
\,=\, 
\frac{1}{d^3} 
\sum_{c \in\{0,1,2,\ldots,d-1\} } \, 
\sum_{b \in\{0,1,2,\ldots,d-1\} } \, 
\sum_{a \in\{0,1,2,\ldots,d-1\} } \, 
\Big \vert Trace
\big [
e^{\frac{2 \pi i}{d}} E_{a,b}^{\otimes} \big ] \Big | ^2 
$$
$$
\,=\, 
\frac{1}{d^3} 
\sum_{c \in\{0,1,2,\ldots,d-1\} } \, 
\sum_{b \in\{0,1,2,\ldots,d-1\} } \, 
\sum_{a \in\{0,1,2,\ldots,d-1\} } \, 
\Big \vert e^{\frac{2 \pi i}{d}} Trace[E_{a,b}^{\otimes} ] \Big | ^2 
$$
$$
\,=\, 
\frac{1}{d^3} 
\sum_{c \in\{0,1,2,\ldots,d-1\} } \, 
\sum_{b \in\{0,1,2,\ldots,d-1\} } \, 
\sum_{a \in\{0,1,2,\ldots,d-1\} } \, 
\Big \vert Trace[E_{a,b}^{\otimes} ] \Big | ^2 
$$
$$
\,=\, 
\frac{1}{d^3} 
\,d\, 
\sum_{b \in\{0,1,2,\ldots,d-1\} } \, 
\sum_{a \in\{0,1,2,\ldots,d-1\} } \, 
\Big \vert   \, d \, \delta_{a,0}  \, \delta_{b,0}  \, \Big | ^2 
\,=\, 
1.
$$ 
Thus we find the group $\mathcal{Q}$ is irreducible. 

The fact that the channel $\mathcal{E}^{\otimes}$ is diagonal in the 
operator basis $\Big \{ E_{a,b}^{\otimes}\Big \}$, coupled with the
equation 
(\ref{E:10}) 
result that $E_{g,h}^{\otimes} \, E_{a,b}^{\otimes} 
\, E_{g,h}^{\otimes^{\dagger}} \,= \, \Omega^{c} \, E_{a,b}^{\otimes}$,
and the equation (\ref{E:11}) result that
$\mathcal{E}\left ( E_{a,b}^{\otimes} \right ) \,=\, \Lambda_{a,b} 
\, E_{a,b}^{\otimes}$, 
allows us to conclude the operators $\Big \{F_{a,b,c}^{\otimes} \Big \}$ 
and the channel $\mathcal{E}^{\otimes}$ commute.
\begin{equation}\label{E:14}
F_{g,h,j}^{\otimes}\, \mathcal{E}\left ( \rho \right ) \, 
F_{g,h,j}^{\otimes^{\dagger}} 
\,=\, 
E_{g,h}^{\otimes}\, \mathcal{E}\left ( \rho \right ) \, 
E_{g,h}^{\otimes^{\dagger}}
\,=\, 
E_{g,h}^{\otimes}\, 
\left ( \, \frac{1}{d} \, \sum_{a,b} \, \alpha_{a,b} \, \Lambda_{a,b} \, 
E_{a,b}^{\otimes}   
\, \right ) \, 
E_{g,h}^{\otimes^{\dagger}}
\end{equation}
$$
\,=\, 
\frac{1}{d} \, \sum_{a,b} \, \alpha_{a,b} \, 
\Lambda_{a,b} \, 
E_{g,h}^{\otimes}\, 
E_{a,b}^{\otimes}\, 
E_{g,h}^{\otimes^{\dagger}}
\,=\,
\mathcal{E}\left ( 
E_{g,h}^{\otimes}\, 
\rho 
E_{g,h}^{\otimes^{\dagger}} 
\right ) \, 
\,=\,
\mathcal{E}\left ( 
F_{g,h,j}^{\otimes}\, 
\rho 
F_{g,h,j}^{\otimes^{\dagger}} 
\right ). 
$$
Note that the product channel analysis in equation (\ref{E:14})
is essentially the same derivation as was done in equation 
(\ref{E:12}) for qudits in the $\hat{X}^a \, \hat{Z}^b$ operator basis.  

This is the key criterion for ensemble achievability.
Since the $\Big \{ F_{a,b,c}^{\otimes}\Big \}$ are
unitary, $F_{g,h,j}^{\otimes}\, 
\rho 
F_{g,h,j}^{\otimes^{\dagger}} $ is a valid density operator. 
Applying any member of $\big \{ F_{a,b,c}^{\otimes}\big \}$ to 
an output optimal ensemble $\{ p_i \,,\, {\widetilde{\rho}_{i}}^{\otimes}\}$
yields an achievable ensemble. Since the 
group $\big \{ F_{a,b,c}^{\otimes}\big \}$
is irreducible, we can  
apply Schur's lemma and conclude the average output 
state  $\widetilde{\Phi}^{\otimes}$ for an optimal ensemble for the product 
channel $\mathcal{E}^{\otimes}$ must equal $\frac{1}{d} \, I_d$.

The remainder of our analysis for diagonal unital qudit
channels 
uses the Schumacher and Westmoreland results summarized in equations
(\ref{E:4}), (\ref{E:5}), (\ref{E:6}) and (\ref{E:7}) in the manner seen
previously, and directly
carries over to the product channel case.
Thus we conclude for the product channel $\mathcal{E}^{\otimes}$,
the HSW channel capacity is
$$
\mathcal{C} \,=\, \log_2(d) \,-\, min_{\rho} \, \mathcal{S}
\left ( \mathcal{E}^{\otimes}(\rho) \right ) 
\,=\,
\sum_{k=1}^{N} \, \log_2(d_k) \,-\, min_{\rho} \, \mathcal{S}
\left ( \mathcal{E}^{\otimes}(\rho) \right ). 
$$

\section{Discussion and Conclusions} 

The HSW channel capacity for single qubit unital channels was 
originally derived in \cite{Ruskai99a} as 
$$
\mathcal{C} \,=\, 1  \,-\, min_{\rho} \, \mathcal{S}
\left ( \mathcal{E}(\rho) \right ). 
$$
This result was extended in \cite{king} to the tensor product
of single qubit unital channels. 
For qubits, it was shown in \cite{rsw} that there always exists 
a special basis in which a qubit unital channel can be written
in diagonal form. A key step in their proof was a homomorphism 
between $SU(d)$ and $SO(d^2-1)$. Such a homomorphism would be necessary 
for the method of proof in \cite{rsw} to carry through to the general
qudit case for $d>2$. 
However this homomorphism only occurs for $d=2$.
Our method 
for deriving the HSW channel capacity depends on the qudit 
unital channel being diagonal, so our method only allows us
to conclude that 
$$
\mathcal{C} \,=\, \log_2(d)  \,-\, min_{\rho} \, \mathcal{S}
\left ( \mathcal{E}(\rho) \right )
$$
holds for diagonal unital channels. 
However, our proof was handcrafted in two key respects. The first 
was the choice of a fixed operator basis, the Generalized Pauli basis,
in which the density matrix expansions were made. There exists 
the possibility that, given a specific channel, a custom operator basis
could be constructed in which the channel $\mathcal{E}$ would 
be diagonal. This in essence is how the 
proof showing any unital qubit channel is diagonal in some operator
basis, was done in \cite{Ruskai99a}. 

The second 
assumption was the explicit manner by which we showed ensemble 
achievability. To summarize, we showed an 
output ensemble was achievable by

1) restricting our
attention to a preordained unitary operator basis consisting of elements
$g \in \mathcal{G}$

and

2) considering only diagonal channels in the basis $\mathcal{G}$.

The result was an algorithm by which we were able
to determine, given an optimal ensemble \oe, if the output ensemble
$\Big \{ p_i ,g \widetilde{\rho}_i g^{-1}\Big \}$  
was achievable for $g \in \mathcal{G}$. 

The possibility remains that, given a channel $\mathcal{E}$, we could
use a technique other 
than that developed in this paper to assure
ensemble achievability across all elements of a group 
$\mathcal{G}$ acting on the channel outputs of an optimal input ensemble.
Again, this is essentially what occurs in the unital qubit channel 
scenario analyzed in \cite{Ruskai99a}. 

As a result, we feel we have ``overconstrained'' the requirements for
our proofs. We conjecture the relation 
$$
\mathcal{C} \,=\, \log_2(d)  \,-\, min_{\rho} \, \mathcal{S}
\left ( \mathcal{E}(\rho) \right )
$$
holds for all unital qudit channels, rather than just those 
unital channels which are diagonal in the Generalized Pauli basis. 

As our final remark, the diagonal unital qudit channel capacity result
extends the connection between the minimum von Neumann 
entropy at the channel output and the HSW channel capacity, which 
had previously been established in the qubit case, 
to a non-empty set of 
channels in any dimension. This implies a more universal connection 
between the minimum von Neumann entropy at the channel output and the 
classical information capacity for that quantum channel than 
had previously 
been shown. 

Furthermore, recall that 
the Holevo quantity $\chi$ utilizes von Neumann entropy to obtain a 
relation for the distinguishability of quantum states.
Hence it is reassuring that von Neumann entropy appears 
explicitly in our
qudit channel capacity result. This is an indicator of consistency 
that reaffirms the fundamental role von Neumann entropy 
appears to play in Quantum Information Science. 

\section{Acknowledgements} 

The author would like to thank Patrick Hayden, Eric Rains and David Bacon
for fruitful discussions during the development of this paper.  

\newpage

\begin{appendix}

\section{The Generalized Pauli Group}

The generalized Pauli operators $\hat{X}$ and $\hat{Z}$ are used 
in our qudit analysis. This section describes some of the properties 
of these operators. Their definitions are
$$
\hat{X} \big | j \big \rangle \;=\; \big | j+1 \, (mod \, d) \big \rangle \qquad 
and \qquad  
\hat{Z} \big | j \big \rangle \;=\; \Omega^j \, \big | j \big \rangle.
$$
The quantity $\Omega \;=\; e^{\frac{2\pi i}{d}}$.  
Note that $\hat{X}^d \;=\; \hat{Z}^d \;=\; I_d$.  
The commutation relation of 
$\hat{X}$ and $\hat{Z}$ follows directly, yielding 
$\hat{Z}\hat{X} \;=\; \Omega \hat{X}\hat{Z}$.  
Using the fact that 
$\big \langle j+1 \big |  \hat{X} \big | j \big \rangle \,=\, 1$, 
taking the Hermitian conjugate of both sides yields 
$\big \langle j \big |  \hat{X}^{\dagger} \big | j+1 
\big \rangle \,=\, 1$, 
allowing us to conclude 
$\hat{X}^{\dagger} \big | j \big \rangle \;=\; 
\big | j-1 \, (mod \, d) \big \rangle$. This in turn implies $\hat{X}$
is unitary, since 
$\hat{X} \hat{X}^{\dagger} \;=\; \hat{X}^{\dagger} \hat{X}\,=\, I_d$.  
Similarly
$\hat{Z}^{\dagger} \big | j \big \rangle \,=\,
\Omega^{-j} \, \big | j \big \rangle$, from 
which it follows that $\hat{Z}$ 
is a unitary operator. 

In our application of Schur's Lemma, we 
use the operator set of $E_{a,b} \,=\, \hat{X}^a \hat{Z}^b$,
where $\{a,b\} \,=\, 0,1,2,\cdots,d-1$. We shall also use
the operators $F_{a,b,c} \,=\, \Omega^c \hat{X}^a \hat{Z}^b$,
where $\{a,b,c\} \,=\, 0,1,2,\cdots,d-1$. 
The operators $E_{a,b}$ and $F_{a,b,c}$ are unitary, 
since the composition of unitary operators is unitary.
Note that $E_{a,b}^{\dagger}\,=\, \hat{Z}^{-b} \hat{X}^{-a}$ and
$F_{a,b,c}^{\dagger} \,=\, \Omega^{-c} E_{a,b}^{\dagger}$.  

We now show that any qudit density operator $\rho$ can be expanded 
as
$$
\rho \;=\; \frac{1}{d} \, \sum_{a,b\in 
\{0,1,2,\cdots,d-1\}} \; \alpha_{a,b} \; \hat{X}^a \; \hat{Z}^b
\;=\; \frac{1}{d} \, \sum_{a,b\in 
\{0,1,2,\cdots,d-1\}} \; \alpha_{a,b} \; E_{a,b},
$$ 
where the $\alpha_{a,b}$ are complex quantities. 
We shall work in the Hilbert-Schmidt operator norm, which for 
qudit operators $A$ and $B$ is defined as
$\langle A , B \rangle \;=\; Trace\big[ A^{\dagger} B \big ]$.
Define
the rescaled operators
$Q_{a,b} \;=\; \frac{E_{a,b} }{\sqrt{d}} \;=\; \frac{\hat{X}^a \hat{Z}^b}{\sqrt{d}}$. The operators $Q_{a,b}$ are a set of $d^2$ 
orthonormal operators in the Hilbert-Schmidt
inner product, as shown below.
\begin{equation}\label{E:15}
\big \langle Q_{a,b} \,,\, Q_{q,r} \big \rangle  \;=\;
\frac{1}{d} \, \big \langle E_{a,b} \,,\, E_{q,r} \big \rangle  \;=\;
\frac{1}{d} \, Trace\big[ E_{a,b}^{\dagger} \, E_{q,r} \big] \;=\; 
\frac{1}{d} \, Trace \big [ 
\hat{Z}^{-b} \hat{X}^{-a} \, 
\hat{X}^{q} \hat{Z}^{r}  
\big  ]
\end{equation}
$$
(By \; the\; cyclic\; nature\; of\; trace) \;=\; 
\frac{1}{d} \, 
Trace \big [ 
\hat{X}^{q-a} \hat{Z}^{r-b}
\big  ]
\;=\; \frac{1}{d} \, \sum_{j=0}^{d-1} \; 
\big \langle j \big | 
\hat{X}^{q-a} \hat{Z}^{r-b}
\big | 
j \big \rangle 
$$
$$
\;=\; \frac{1}{d} \, \sum_{j=0}^{d-1} \; \Omega^{(r-b)j } \,  
\big \langle j \big | 
\hat{X}^{q-a} 
\big | 
j \big \rangle 
\;=\; \frac{1}{d} \, \sum_{j=0}^{d-1} \; \Omega^{(r-b)j } \,  
\big \langle j \big | 
j+q-a \, (mod \, d)  \big \rangle 
$$
$$ 
\;=\; \frac{1}{d} \, \delta_{a,q} \; \sum_{j=0}^{d-1} \; \Omega^{(r-b)j }   
\;=\; \frac{1}{d} \, d \; \delta_{a,q} \; \delta_{b,r} 
\;=\; \delta_{a,q} \; \delta_{b,r}. 
$$ 
Here $\delta_{\alpha,\beta} $ is the Kronecker delta function.  
Recall any qudit density operator $\rho$ can be written as 
$$
\rho \;=\; \sum_{a=0}^{d-1} \, \sum_{b=0}^{d-1} \; \beta_{a,b} \, 
\big | a  \big \rangle  \big  \langle b \big  |,
$$
where the $\beta_{a,b}$ are complex quantities. 
We shall show that 
$\big | a  \big \rangle  \big  \langle b \big  |$
may be written as 
$\big | a  \big \rangle  \big  \langle b \big  | \;=\; 
\sum_{r=0}^{d-1} \, \sum_{s=0}^{d-1} \; \zeta_{r,s} \, Q_{r,s}$,
where the $\zeta_{r,s}$ are complex quantities.
Rescaling the $\zeta_{r,s}$, we will conclude 
that $\rho$ may be written as
$$
\rho\;=\; \sum_{a=0}^{d-1} \, \sum_{b=0}^{d-1} \; \alpha_{a,b} \, E_{a,b}.
$$ 

To begin, write $Q_{r,s}$ as 
$$
Q_{r,s} \;=\; \frac{1}{\sqrt{d}} \, \sum_{j=0}^{d-1} \,
\Omega^{js} \, \big | j+r\big \rangle \big \langle j \big |.
$$
Define 
$\zeta_{r,s}$ as \cite{preskill}
\begin{equation}\label{E:a}
\zeta_{a,b} \;=\; 
Trace\Bigg[ Q_{r,s}^{\dagger} \big | a \big \rangle \big \langle b \big |
\Bigg ] \;=\; 
\frac{1}{\sqrt{d}} \, Trace\Bigg[ \sum_{j=0}^{d-1} \, \Omega^{-js} 
\, \big | j\big \rangle \big \langle j+r \big |
a \big \rangle \big \langle b \big | \Bigg ]
\end{equation}
$$
 \;=\;
\left (Do \quad the \quad Trace \quad in \quad the\quad  basis \quad 
\Big \{ \big | i \big \rangle \Big \} \right )
\longrightarrow
\frac{1}{\sqrt{d}} \, 
\sum_{i=0}^{d-1} \, \sum_{j=0}^{d-1} \, \Omega^{-js} 
\, \big \langle i \big | j\big \rangle \big \langle j+r \big |
a \big \rangle \big \langle b \big |  i \big \rangle  
$$
$$
\;=\;
\frac{1}{\sqrt{d}} \, 
\sum_{i=0}^{d-1} \, \sum_{j=0}^{d-1} \, \Omega^{-js} \, 
\delta_{b,i} \, 
\delta_{j+r,a} \, 
\delta_{i,j} \;=\; 
\frac{1}{\sqrt{d}} \, 
\sum_{j=0}^{d-1} \, \Omega^{-js} \, 
\delta_{j,b} \, 
\delta_{j+r,a} \, 
\;=\; 
\frac{1}{\sqrt{d}} \, 
\Omega^{-bs} \, 
\delta_{a,b+r} ,
$$
where $\delta$
is the Kronecker delta function.  

Consider the operator $L\,=\, \ab$,
and the corresponding complex coefficients $ \xi_{r,s} \,=\,  
\big \langle Q_{r,s} , L \big \rangle  \,=\, 
Trace\Big[ Q_{r,s}^{\dagger} \big | a \big \rangle \big \langle b \big |
\Big ] $. We would like to expand $L$ as 
$L \,=\, \sum_{r,s} \, \big \langle Q_{r,s} \,,\, L \big \rangle \, Q_{r,s}
\,=\, \sum_{r,s} \, \xi_{r,s} \, Q_{r,s}$.  
Note that $\| L \| \,=\, \sqrt{\big \langle L , L \big \rangle } \,=\, 1$. 
Using the result of equation (\ref{E:a}), we can conclude that 
$$
\sum_{r} \, \sum_{s} \, \left | \xi_{r,s} \right |^2 \,=\, 
\sum_{r} \, \sum_{s} \, \left| \frac{1}{\sqrt{d}} \, \Omega^{-bs} \, 
\delta_{a,b+r} \right |^2 \,=\,  
\frac{1}{d} \, \sum_{r} \, \sum_{s} \, \left | \delta_{a,b+r} \right |^2 \,=\, 
\frac{1}{d} \, d \, \sum_{r} \, \left | \delta_{a,b+r} \right |^2 \,=\, 1.
$$
Thus 
$ \sum_{r} \, \sum_{s} \, | \xi_{r,s} |^2 \,=\,  1 \, =\, \| L \|^2$.
This fact for arbitrary $a$ and $b$ in $\ab$ allows us to conclude the 
$Q_{r,s}$ form a complete, orthonormal basis for the $L$'s, and 
we can expand $L$ in terms of the $Q_{r,s}\; \forall \; a,b$\cite{halmos}.
Thus the expansion 
$
\ab 
 \,=\, \sum_{r,s} \, 
\big \langle Q_{r,s} \,,\, L \big \rangle \, Q_{r,s}$ 
holds $\forall \; a,b$.  
This leads to an expansion for the qudit density operator $\rho$. 
\begin{equation}
\rho \,=\, \sum_{a=0}^{d-1} \, \sum_{b=0}^{d-1} \; \beta_{a,b} \, 
\big | a  \big \rangle  \big  \langle b \big  |
\,=\, 
\sum_{a=0}^{d-1} \, \sum_{b=0}^{d-1} \; \beta_{a,b} \, 
\sum_{r=0}^{d-1} \, \sum_{s=0}^{d-1} \,
\Big \langle \, Q_{r,s} \,,\, \left (\ab \right )  \,  \Big \rangle \, Q_{r,s} 
\end{equation}
$$
\,=\, 
\sum_{r=0}^{d-1} \, \sum_{s=0}^{d-1} \,
\sum_{a=0}^{d-1} \, \sum_{b=0}^{d-1} \, \beta_{a,b} \, 
\Big \langle \, Q_{r,s} \,,\, \left ( \ab \right )  \,  \Big \rangle \, Q_{r,s} 
\,=\, 
\sum_{r=0}^{d-1} \, \sum_{s=0}^{d-1} \,
\left  \langle \, Q_{r,s} \,,\, 
\left ( \, \sum_{a=0}^{d-1} \, \sum_{b=0}^{d-1} \,
 \beta_{a,b} \ab \,  \right )  \,  \right \rangle 
\, Q_{r,s} 
$$
$$
\,=\, 
\sum_{r=0}^{d-1} \, \sum_{s=0}^{d-1} \,
\Big \langle \, Q_{r,s} \,,\, 
\rho  \,  \Big \rangle 
\, Q_{r,s} 
\,=\, 
\sum_{r=0}^{d-1} \, \sum_{s=0}^{d-1} \,
\frac{\alpha_{r,s}}{\sqrt{d}} \, Q_{r,s} 
\,=\, 
\sum_{r=0}^{d-1} \, \sum_{s=0}^{d-1} \,
\frac{\alpha_{r,s}}{\sqrt{d}} \, \frac{E_{r,s}}{\sqrt{d}} 
\,=\, 
\frac{1}{d} \, \sum_{r=0}^{d-1} \, \sum_{s=0}^{d-1} \,
\alpha_{r,s} \, E_{r,s} 
$$ 
$$
\quad 
where \quad 
\frac{\alpha_{r,s}}{\sqrt{d}} \,=\, 
\big \langle \, Q_{r,s} \,,\, \rho  \,  \big \rangle
\quad or \quad equivalently \quad \alpha_{r,s}  
\,=\, \big \langle \, E_{r,s} \,,\, \rho  \,  \big \rangle.$$
The linearity of the inner product in the second argument 
was used to move the sum over the indices 
$a$ and $b$ inside the inner product. 

To obtain the final form of the expansion for 
the qudit operator $\rho$ we shall use, 
note that $E_{0,0} \,=\, I_d$. Our result above, 
$
\big \langle E_{a,b} \,,\, E_{q,r} \big \rangle  \,=\,
Trace\big[ E_{a,b}^{\dagger} \, E_{q,r} \big] \,=\,
d \, \delta_{a,q} \; \delta_{b,r} 
$, tells us that 
$Trace(E_{a,b}) \;=\; d \, \delta_{a,0} \, \delta_{b,0}$. 
Thus of the $d^2$ possible $E_{a,b}$, only 
$E_{0,0}$ has nonzero $Trace$. 
The trace condition $Trace(\rho) \,=\, 1$ allows us to conclude
$\alpha_{0,0} \,=\, 1$. 
Using this, 
let $\Upsilon$ denote the set of $d^2 \,-\,1$ elements 
$a,b \;\in\; \{0,1,2,\cdots,d-1\}$  {\it with the exception that 
$a$ and $b$ cannot both be zero.} Then we may write the qudit 
density matrix $\rho$ as 
$\rho \,=\, \frac{1}{d} \, \left ( I_d \;+\; \sum_{(a,b) \in \Upsilon } \; 
\alpha_{a,b} \; E_{a,b} \right ) $ 
with  
$\alpha_{a,b}  
\,=\, \big \langle \, E_{a,b} \,,\, \rho  \,  \big \rangle\,=\,
Trace\big[ E_{a,b}^{\dagger} \rho \big ]$.

In the expansion of $\rho$ above, there are $2d^2 -2$ real,
independent degrees of freedom in the set of coefficients $\alpha_{a,b}$.
However, in the density operator $\rho$, there are only $d^2 -1$ real, 
independent degrees of freedom. Hence there are constraint relations between
the $\alpha_{a,b}$. These constraints arise from 
the Hermitian nature of $\rho$. Note that $E_{a,b}^{\dagger} \,=\, 
\left ( \hat{X}^{a} \hat{Z}^{b} \right ) ^{\dagger} \,=\,
\hat{Z}^{-b} \hat{X}^{-a} \,=\, 
\Omega^{d-b)(d-a) } \hat{X}^{d-a} \hat{Z}^{d-b}
\,=\, 
\Omega^{d-b)(d-a) } E_{d-a,d-b}$. 
Consideration of 
$\rho^{\dagger} \,=\, \rho$ then implies 
$$
\frac{1}{d} \, \left ( I_d \;+\; \sum_{(a,b) \in \Upsilon } \; 
\alpha_{a,b} \; E_{a,b} \right ) 
\,=\, \frac{1}{d} \, \left ( I_d \;+\; \sum_{(a,b) \in \Upsilon } \; 
\alpha_{a,b}^{*} \; E_{a,b}^{\dagger} \right )  
$$
$$
\,=\, \frac{1}{d} \, \left ( I_d \;+\; \sum_{(a,b) \in \Upsilon } \; 
\alpha_{a,b}^{*} \; \Omega^{(d-a)(d-b)} E_{d-a,d-b} \right ) 
$$ 
or
$\alpha_{d-a,d-b} \,=\, \alpha_{a,b}^{*} \Omega^{(d-a)(d-b) }$. 
Here * indicates complex conjugation, and index arithmetic is modulo $d$. 

For example, for qubits, $d=2$, and $\Omega \,=\, e^{\frac{2\pi i}{2} } \,=\, 
e^{\pi i} \,=\, -1$. Applying the constraint equation above  
leads to  $\alpha_{0,1}^{*} \Omega^{(2-0)(2-1)} \,=\,  \alpha_{2-0,2-1}$ 
or $\alpha_{0,1}^{*}  \,=\,  \alpha_{0,1}$, implying the 
coefficient of $E_{0,1} \,=\, \hat{Z}$ must be real.  
Similarly, 
$\alpha_{1,0}^{*} \Omega^{(2-1)(2-0)} \,=\,  \alpha_{2-1,2-0}$
or 
$\alpha_{1,0}^{*}  \,=\,  \alpha_{1,0}$, implying the 
coefficient of $E_{1,0} \,=\, \hat{X}$ must be real.  
Lastly, 
$\alpha_{1,1}^{*} \Omega^{(2-1)(2-1)} \,=\,  \alpha_{2-1,2-1}$
or 
$- \alpha_{1,1}^{*}  \,=\,  \alpha_{1,1}$, implying the 
coefficient of $E_{1,1} \,=\, \hat{X} \hat{Z}$ must be pure imaginary.  
Note that 
$\hat{X} \,=\, \sigma_x$, $\hat{X}  \hat{Z} \,=\, - i \sigma_y$, 
and 
$\hat{Z} \,=\, \sigma_z$. 
Hence we have reproduced the Bloch Sphere representation 
for qubits, 
$\rho \,=\, \frac{1}{2} \left ( I_2 \,
+\, \alpha_{1,0} \hat{X} 
+\, \alpha_{1,1} \hat{X} \hat{Z}  
+\, \alpha_{0,1} \hat{Z} \right )  
\,=\, 
\frac{1}{2} \left ( I_2 \,
+\, w_x  \sigma_x
+\, i w_y  \left ( -i\sigma_y \right )  
+\, w_z \sigma_z \right ) $,
with the $w_k$ real. For qubits,
we end up with $3\,=\, d^2-1$ real independent 
parameters, and not $2d^2 -2\,=\, 6$. The constraint equations for the 
$\alpha_{a,b}$ eliminated three real degrees of freedom. In general, 
the constraint equations will eliminate $d^2-1$ real extra degrees of freedom,
leaving $d^2 -1$ actual real parameters.

\end{appendix}

\end{document}